\documentclass[aps,prl,reprint,showpacs,footinbib,amsmath,amssymb,amsfonts,floatfix,superscriptaddress]{revtex4-1}

\usepackage{graphicx}
\usepackage{epsfig}
\usepackage{color}

\begin{document}

\title{Composition variation and underdamped mechanics \\ near membrane proteins and coats}



\author{S. Alex Rautu}
\affiliation{Department of Physics, University of Warwick, Coventry, CV4 7AL, United Kingdom}

\author{George Rowlands}
\affiliation{Department of Physics, University of Warwick, Coventry, CV4 7AL, United Kingdom}

\author{Matthew S. Turner}
\affiliation{Department of Physics, University of Warwick, Coventry, CV4 7AL, United Kingdom}
\affiliation{Centre for Complexity Science, University of Warwick, Coventry CV4 7AL, United Kingdom}


\date{\today}

\begin{abstract}

We study the effect of transmembrane proteins on the shape, composition and thermodynamic stability of the surrounding membrane. When the coupling between membrane composition and curvature is strong enough the nearby membrane composition and shape both undergo a transition from over-damped to under-damped spatial variation, well before the membrane becomes unstable in the bulk. This transition is associated with a change in the sign of the thermodynamic energy and hence favors the early stages of coat assembly necessary for vesiculation (budding) and may suppress the activity of
mechanosensitive membrane channels and transporters. Our results suggest an approach to obtain physical parameters of the membrane that
are otherwise difficult to measure.

\end{abstract}



\maketitle

Biological membranes are crucial to the structure and function of living cells \cite{molecular-bio}. Transmembrane proteins essential for transport, adhesion and signalling are embedded in membranes \cite{Singer1972, Engelman2005} consisting of a mixture of lipids and other amphipathic components. The interaction with the adjacent lipid molecules is known to regulate the function of membrane proteins \cite{Suchyna2004, Moe2005, Krepkiy2009, Milescu2009}. Here, we are primarily interested in the non-specific lipid-protein interactions that arise from the coupling of their hydrophobic regions \cite{Lee2004, Mitra2004, Dowhan2004, Nyholm2007, Andersen2007, Phillips2009, Lundbaek2010}, although we can also allow for selective enrichment of membrane component(s) near the protein. We employ a continuum theory in which small deformations of the lipid environment near a rigid inclusion can be described by a number of local field variables, such as the profile of the mid-plane of the bilayer, its composition and membrane thickness \cite{Canham1970, Huang1986, Aranda-Espinoza1996, Nielsen1998, Mondal2014, Sens2000, Fournier1999, Weikl1998, Wiggins2004, Wiggins2005, Dan1993, Goulian1993, Kim1998, Turner1999, Markin2004, Ursell2007, Haselwandter2012, Haselwandter2013, leibler1, leibler2, Andelman1992, Helfrich1973}. Furthermore, the free-energy cost associated with thickness deformation is completely decoupled at lowest order \cite{Fournier1999}, and it can be independently analyzed although we do not do so here. 

\begin{figure}[ht]\includegraphics[width=\columnwidth]{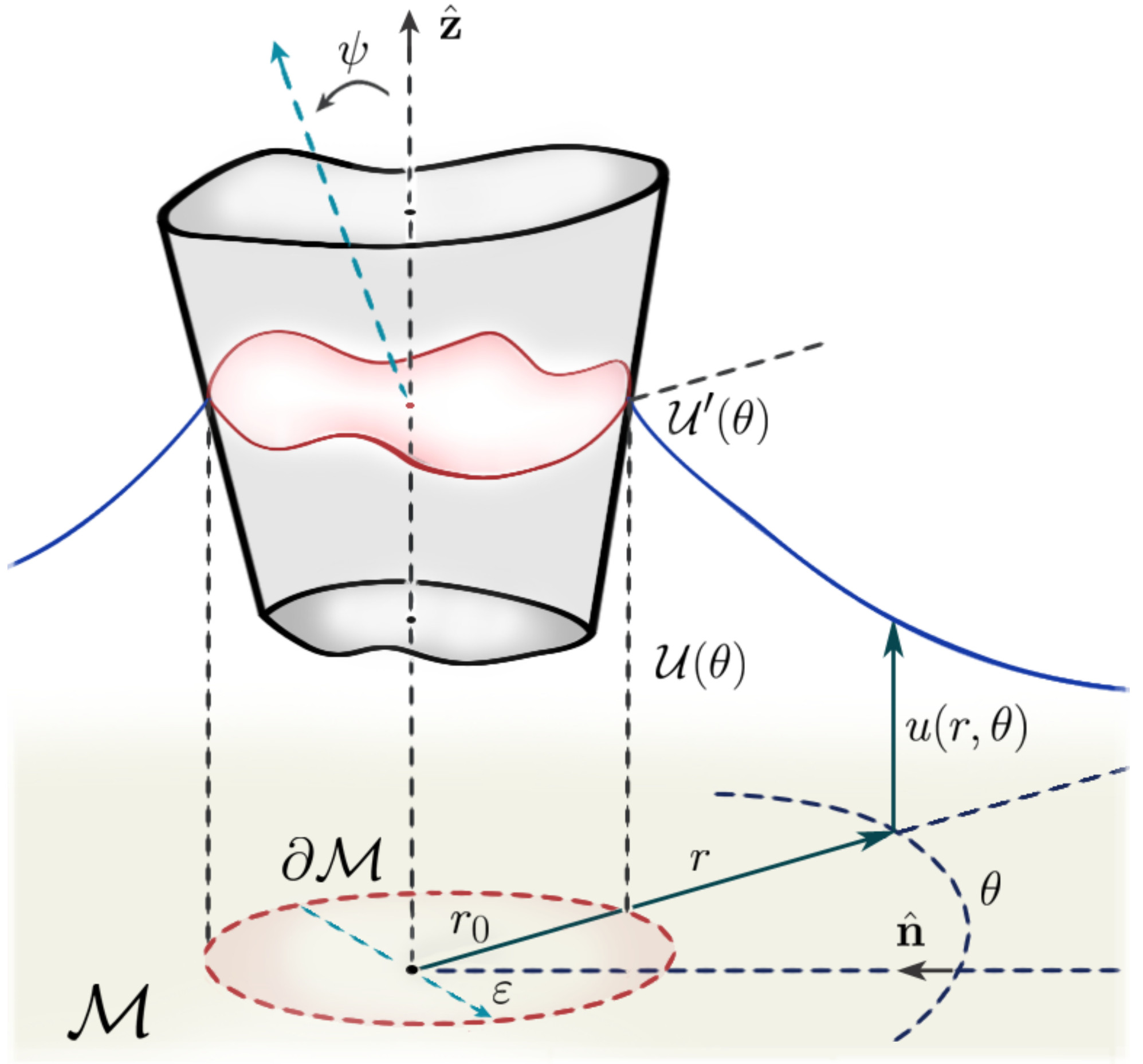}\caption{\label{fig:1} 
(color on-line) Sketch of a membrane inclusion showing the mid-plane of the bilayer (blue line) at height $u(r,\theta)$. The surface variation of the rigid inclusion in the $\mathbf{\hat{z}}$ direction is coarse-grained out so that the geometry is defined by its radius $r_0$ and two functions describing the height $\mathcal{U}(\theta)$ and contact angle $\mathcal{U}'(\theta)$ of the hydrophobic belt. These parameterize the protein-membrane interface (red line), where $u(r_0, \theta)= \mathcal{U}(\theta)$ and $\mathbf{\hat{n}}\cdot \nabla u(r_0, \theta) = \mathcal{U}'(\theta)$, with $\mathbf{\hat{n}}$ as the inward unit normal vector. We require both the normal force and the torque on the inclusion to vanish. The latter can lead to an equilibrium tilt angle $\psi$ about the axis labeled by $\varepsilon$.}
\end{figure}

We allow for selective enrichment/depletion of curvature sensitive inclusions in the vicinity of a membrane protein or, equivalently, lipid asymmetry between leaflets that is characterized by a local spontaneous curvature, the preferred mean curvature in the absence of any mechanical stresses on the membrane \cite{leibler1,leibler2, Andelman1992, Helfrich1973, Seifert1997, Israelachvili1976, Gruner1989, McMahon2005, Callan-Jones2011}. This local variation may be relatively large near a membrane protein if its geometry is such that it bends or deforms the surrounding membrane (see Fig~\ref{fig:1}). Our approach leads to a real-space description of the membrane around an inclusion of arbitrary symmetry. 

We consider a two-component membrane in which the local compositional asymmetry between the different layers and/or the density of curvature-sensitive inclusions is phenomenologically coupled to the local mean curvature of the membrane \cite{leibler1, leibler2}.  When the compositional variation is weak and the membrane displacement is small, the free-energy can be written as a Landau-Ginzburg expansion \cite{leibler1, leibler2, Andelman1992, SunilKumar1999, Schick2012, Seifert1993}, \begin{equation}\label{eqn:F-phi}
    \mathcal{F}_{\varphi} = \frac{1}{2} \int_{\mathcal{M}} \left[a\,\varphi^2 + b\,(\nabla \varphi)^2 + 
    2\,c\,\varphi\left(\nabla^2 u\right)\right] \mathrm{d}^2\mathbf{r},
\end{equation} where only the lowest-order terms are retained and $a$, $b$ and $c$ are phenomenological constants. The scalar fields $\varphi(\mathbf{r})$ and $u(\mathbf{r})$ are the local composition difference (as an area fraction) and bilayer mid-plane height, respectively, see Fig~\ref{fig:1}. Both deformation fields are described within a Monge representation, which allows us to write the free-energy associated with mid-plane deformation as 
\begin{equation}\label{eqn:F-u}
    \mathcal{F}_{u} = \frac{1}{2}\int_{\mathcal{M}} \left[\sigma\,(\nabla u)^2 + \kappa\,(\nabla^2 u)^2\right] \mathrm{d}^2\mathbf{r},
\end{equation} where  $\sigma$ and $\kappa$ are the surface tension and bending rigidity of the membrane, respectively  \cite{statmech}.

We now seek the ground state of the membrane and neglect fluctuations throughout. The membrane shape $u(\mathbf{r})$ and its compositional field $\varphi(\mathbf{r})$ can then be computed exactly by minimizing the free-energy functional, $\mathcal{F} = \mathcal{F}_{u}+\mathcal{F}_{\varphi}$, leading to the Euler-Lagrange equations: 
\begin{gather}
    \label{eqn:EL1}\nabla^2 u = (\nabla^2 - \beta^2)\,\phi, \\
    \label{eqn:EL2}\nabla^2(\nabla^2 - \alpha^2)\,u +\gamma^2\,\nabla^2\phi=0,
\end{gather} where $\phi(\mathbf{r}) = (b/c)\,\varphi(\mathbf{r})$ and the coefficients $\alpha = \sqrt{\sigma/\kappa\,}$, $\beta = \sqrt{a/b\,}$ and $\gamma = c\,/\sqrt{\kappa b\,}$ represent the relevant inverse length scales of the model \footnote{The sign choice of $\gamma$ is simply a convention. The fields $u$ and $\phi$ are invariant under a sign change in $\gamma$, itself related to the convention of a direction for ``up'' and whether one refers to the enrichment of a component that couples to positive curvature (depending on one's choice for ``up'') or the depletion of one that couples to negative curvature.}. By combining (\ref{eqn:EL1}) and (\ref{eqn:EL2}), a single equation for $\phi(\mathbf{r})$ can be obtained \footnote{See Supplemental Material, which includes Refs. \cite{Riley2006,Byron1992,Gradshteyn2000}.}: \begin{equation}\label{eqn:phi-diff}
    \,(\nabla^2 - k_+^2)(\nabla^2 - k_-^2)\,\phi=0,
\end{equation} where $k_\pm$ is given by \begin{equation}\label{eqn:k_pm}
    k_\pm = \frac{1}{2}\left[\sqrt{(\alpha+\beta)^2 - \gamma^2}
    \pm\sqrt{(\alpha-\beta)^2 - \gamma^2} \,\right].
\end{equation} By separation of variables, a solution to equation (\ref{eqn:phi-diff}) that vanishes in the far-field limit can be found to be \begin{equation}\label{eqn:phi-sol}
    \phi(r, \theta) = \phi_{+}(r, \theta) + \phi_{-}(r, \theta),
\end{equation} where $r$ and $\theta$ are the usual polar coordinates, as illustrated in Fig~\ref{fig:1}, and $\phi_\pm$ is defined by \begin{equation}\label{eqn:phi-series}
    \phi_\pm(r, \theta) = \frac{k_\pm^2}{k^2_\pm - \beta^2}\sum_{n=0}^{\infty} \mathcal{V}^{\pm}_{n}(\theta)\,K_n(k_\pm r),
\end{equation} where $K_n$ are the modified Bessel functions of the second kind of order $n$, and $\mathcal{V}^{\pm}_{n}(\theta) = \mathcal{A}^{\pm}_{n}\cos(n\theta)+\mathcal{B}^{\pm}_{n}\sin(n\theta)$, with $\mathcal{A}^{\pm}_n$ and $\mathcal{B}^{\pm}_n$ arbitrary constants. From this we obtain the membrane shape through Eq~(\ref{eqn:EL1}), which yields  \begin{equation}\label{eqn:u-sol}
    u(r, \theta) = u_{+}(r, \theta) + u_{-}(r, \theta) + u_{h}(r, \theta),
\end{equation} where the solutions that diverge at infinity are excluded. Here, $u_h(r, \theta)$ is the homogeneous solution of (\ref{eqn:EL1}), namely 
\begin{equation}\label{eqn:u-h}
    u_h(r, \theta) = \sum_{n=0}^{\infty} \mathcal{W}_n(\theta)\,r^{-n},
\end{equation} where $\mathcal{W}_n(\theta) = \mathcal{X}_{n}\cos(n\theta) + \mathcal{Y}_{n}\sin(n\theta)$, with $\mathcal{X}_n$ and $\mathcal{Y}_n$ some constants. The remaining two terms in (\ref{eqn:u-sol}) are the inhomogeneous solutions, which are found to be \begin{equation}\label{eqn:u-series}u_\pm(r, \theta) = \sum_{n=0}^{\infty} \mathcal{V}^{\pm}_{n}(\theta)\,K_n(k_\pm r).\end{equation} 

The angular functions $\mathcal{V}^{\pm}_{n}(\theta)$ and $\mathcal{W}_{n}(\theta)$ are determined by the boundary conditions at the interface $\partial\mathcal{M}$, located at a distance $r_0$ from the symmetry axis. These are specified by the height $\mathcal{U}(\theta)$ and contact angle $\mathcal{U}'(\theta)$ at which the mid-plane of the bilayer meets the inclusion (see Fig~\ref{fig:1}). This choice is motivated by assuming a strong coupling between the transmembrane domain of the inclusion and the membrane hydrophobic core. Also, the normal derivative of $\phi$ is chosen to vanish on $\partial\mathcal{M}$, which is used to obtain an unique solution \footnote{Although Dirichlet boundary conditions could be used as well, our choice gives the ground state solution in the absence of any constraints on the composition asymmetry field $\phi$ at the boundary; see Ref. \cite{Note2} for more details}. 

\begin{figure}[t]\includegraphics[width=0.85\columnwidth]{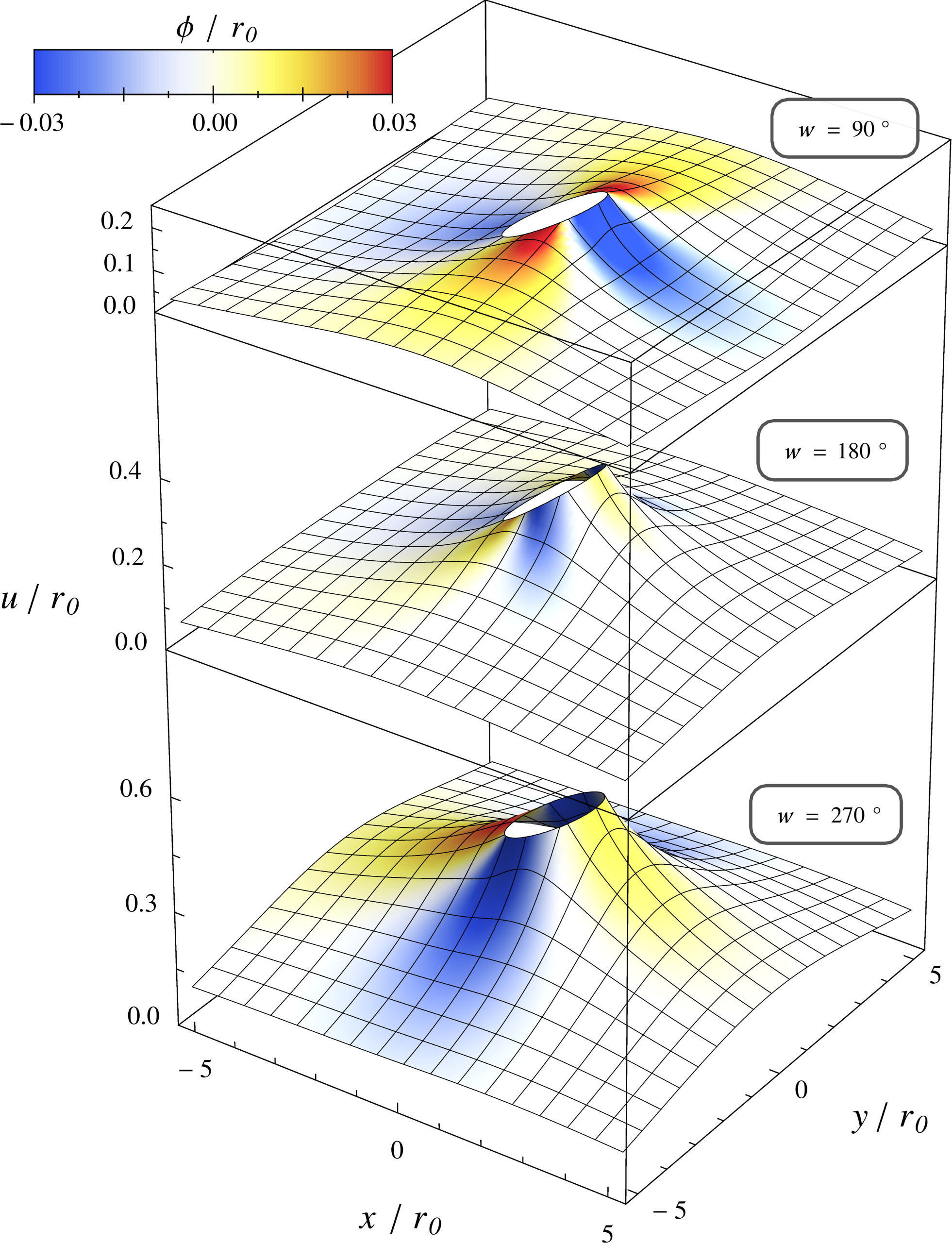}\caption{\label{fig:2} 
(color on-line) Membrane profiles induced by an asymmetrical inclusion, where the contact angle is given by $\mathcal{U}'(\theta) = 15^{\circ}$ if $|\theta|<w/2$ and 0 otherwise, with $\theta$ measured from the $x-$axis (see text). The membrane parameters are here $\alpha\,r_0 = 0.1$, $\beta\,r_0 = 1.0$ and $\gamma\,r_0 = 0.5$, with $r_0$ the radius of the inclusion (not depicted). The compositional asymmetry $\phi(\mathbf{r})$ is shown as the color-map of the surface plots.}\end{figure}

This methodology allows us to compute exactly the lowest order estimates to the membrane profile, its local phase behavior, and the total deformation energy, given an arbitrary model for the shape of the inclusion, through $\mathcal{U}(\theta)$ and $\mathcal{U}'(\theta)$, i.e.~a general solution to the problem. This makes contact with experiments that might measure membrane shape (cryo TEM \cite{Shimada2007} or perhaps TIRF microscopy) and composition (NMR \cite{Judge2011} or FRET \cite{Loura2011}). First, we consider a simple illustrative example in which the height, $\mathcal{U}(\theta)$, is chosen to be a constant $z_0$, while the contact angle has a non-zero value only within an angular interval $w$, see Fig~\ref{fig:2}. This corresponds to a rigid inclusion that induces a local mid-plane deformation only within a specific region along its hydrophobic belt, with the remaining part preferring a flat membrane. The Connolly surface of a leucine transporter, LeuT, exhibits similar features \cite{Yamashita2005, Singh2008}. The height $z_0$ is not entirely arbitrary, being set by the overall balance of normal forces. Similarly, the condition of torque balance leads to a tilt of the inclusion (see \cite{Note2} for details), as illustrated in Fig 1. Typical solutions due to such an asymmetrically-shaped inclusion that exerts no net torque are shown in Fig~\ref{fig:2} for physiologically reasonable values of $\alpha$, $\beta$, and $\gamma$ \footnote{Typically the membrane correlation length might be $\alpha^{-1}\sim 10$ $r_0$ \cite{Sheetz1996}.  Eq.~(\ref{eqn:F-phi}) can be associated with a typical interfacial width for a strongly segregated system of a few nm, say $\sqrt{b/|a|}\sim r_0$, and a line tension of about a pN ($\sqrt{b|a|}\sim$ pN) \cite{Tian2007}, which combine to give $b\sim k_BT$. Sorting of strongly curvature-coupling components into membrane tubes gives an upper bound of $c\sim\kappa/(2.5\,r_0)$ and hence a value for $\gamma \sim 2/r_0$ \cite{Aimon2014}, indicating that the all regimes discussed in the main text are accessible. While the use of a phase separated system to estimate the parameters $a$ and $b$ is questionable, given that they are motivated within a model in which the system remains essentially one phase, with $\phi$ very small,  we are reassured that the primary limit on the accessibility of the underdamped regime is that $\beta$ is not too large. Given that spontaneous phase separation is often seen on biological membranes there would seem to be no lower limit on a reasonable magnitude for $a$ and hence $\beta$.}.  The induced $\phi(\mathbf{r})$ shows a rich variation as the angle $w$ is varied between $0$ and $2\pi$, which correspond to inclusions with a cylindrical and a conical shape, respectively.

\begin{figure}[t]\includegraphics[width=\columnwidth]{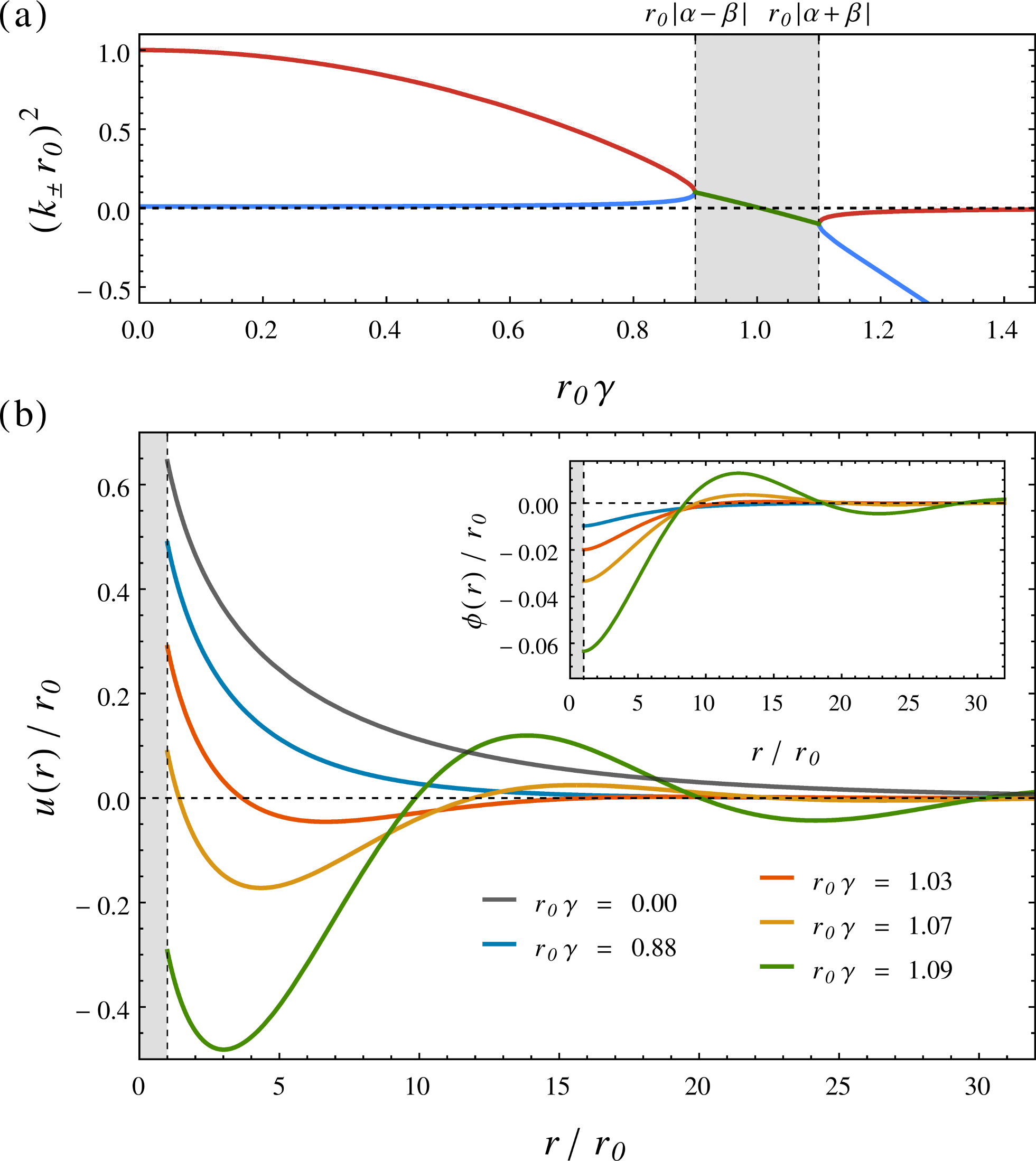}\caption{\label{fig:3} 
(color on-line) (a) Plot of $k^2_{\pm}$ from Eq~(\ref{eqn:k_pm}) against the coupling constant $\gamma$, with $\alpha\,r_0 = 0.1$, $\beta\,r_0 = 1.0$ and $r_0$ the inclusion radius. This illustrates that both $k_{+}$ (red line) and $k_{-}$ (blue line) are real for $\gamma < |\alpha - \beta|$, and purely imaginary when $\gamma > \alpha + \beta$. The grey shaded area illustrates the region where $k^2_{\pm}$ are complex, while the green line shows the real part only. The domain given by $\gamma > \alpha + \beta$ corresponds to Leibler's unstable regime \cite{leibler1, leibler2}. (b) Radial profiles of the bilayer mid-plane $u(r)$ and the compositional asymmetry $\phi(r)$ induced by a conical inclusion, with a modest contact angle of $15^{\circ}$, for different values of the coupling constant $\gamma$, where $\alpha$ and $\beta$ have the same values as those in panel (a).}\end{figure}

To better understand the role of the coupling constant $\gamma$, we consider symmetric conical inclusions ($w=2\pi$) in what follows, noting that the transition from over- to under-damped variation also appears for rigid inclusions with other (or no) symmetry. For values of $\gamma$ less than $\gamma_d = |\alpha-\beta|$, the solutions are found to be monotonically decaying, see Fig~\ref{fig:3}(b). However, as $\gamma$ is increased above this point, the solutions show an underdamped behavior, with the membrane displacement decaying to zero for large distances. 
The magnitude of this amplitude becomes large as $\gamma$ approaches $\gamma_c = \alpha+\beta$, suggesting the presence of an instability. In fact $\gamma>\gamma_c$, where $k^2_\pm<0$ as shown in Fig~\ref{fig:3}(a), corresponds to Leibler's criterion for curvature-induced instabilities in (bulk) membranes \cite{leibler1, leibler2}. The point $\gamma=\gamma_d$ instead corresponds to a critically damped system, separating the real and complex domain of $k_\pm$. The solutions are thermodynamically stable on either side of this boundary. When $\gamma_d<\gamma<\gamma_c$, the decay rate $\lambda$ of the membrane undulations and its wave number $\omega$ can be determined by approximating $K_n (\rho) \approx e^{-\rho}\left(\pi \rho\,/\,2\right)^{-1/2}$ for $\rho\gg n$ in Eq~(\ref{eqn:u-sol}) \cite{Abramowitz1965}, i.e.
\begin{equation}\label{eqn:u-simpl} u(r)\sim \frac{e^{-\lambda(r-r_0)}}{\sqrt{r/r_0}}\,\cos\left[\,\omega\left(r-r_0\right) + \vartheta\,\right],\end{equation} where $\vartheta$ is a phase angle that only depends on $\alpha$, $\beta$ and $\gamma$. Here, $\lambda$ and $\omega$ are given by the real and imaginary parts of (\ref{eqn:k_pm}), respectively. 
Thus, we find that the wavelength of the pre-critical undulations diverges as we approach $\gamma = \gamma_d$, and the decay length diverges for $\gamma = \gamma_c$, which signals the presence of a bulk membrane instability. Physically and mathematically distinct underdamped solutions have also been found in studies of membrane thickness mismatch without any compositional field that couples to mid-plane curvature \cite{Aranda-Espinoza1996}.

While $\alpha$ can be measured through various experimental techniques \cite{Evans1990, Bo1989, Kummrow1991, Pecreaux2004}, the parameters $\beta$ and $\gamma$ are more elusive \cite{Note3}. Our analysis suggests a possible method to measure them, e.g.~by tuning the system to lie near the instability threshold $\gamma\lesssim\gamma_c$. Here, the amplitude of the undulations are large and long-ranged, and $\gamma$ and $\beta$ can be inferred by comparison with (\ref{eqn:u-sol}) or (\ref{eqn:u-simpl}). This tuning might be achieved by controlling the surface tension, e.g.~using a micropipette aspiration technique \cite{Evans1990}, so as to approach the critical tension $\sigma_c=\kappa(\gamma -\beta)^2$, although the presence of thermal fluctuations may mean that some averaging will be required to resolve the ground state, particularly far away from the membrane inclusion. This illustrates the predictive power of our model.

\begin{figure}[t]\includegraphics[width=\columnwidth]{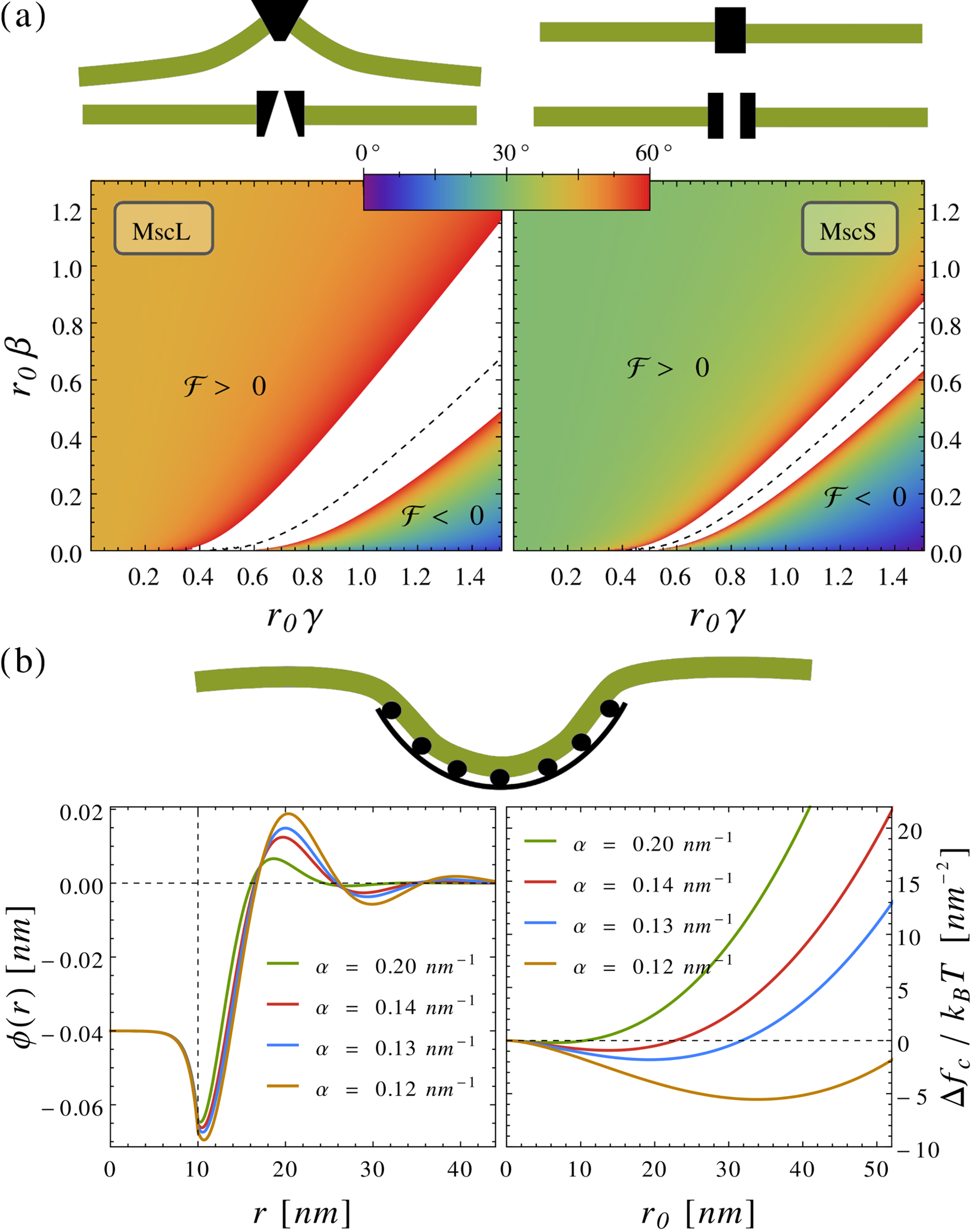}\caption{\label{fig:4} 
(color on-line) (a) The top sketches, with the bilayer membrane represented by a thick green line, show two idealized schemes for channel gating: the gating-by-tilt model (left); and the dilational gating model (right). The figures below this show the angle for gating-by tilt that would account for the absolute value of the \textit{entire} conformational energy change $\mathcal{F}$ measured for MscL and MscS \cite{Chiang2004, Sukharev2002}. The dashed line indicates $\mathcal{F} = 0$ separating two domains where the membrane acts to open ($\mathcal{F} > 0$), or close ($\mathcal{F} < 0$), the gating channel. The uncolored region is not shown as it corresponds to angles greater than $60^\circ$, which are probably unphysical and where our perturbative approach anyway breaks down. 
(b) The top sketch shows a membrane deformed by the assembly of a protein coat such as clathrin or a viral coat protein. The graphs below this show the radial compositional field $\phi(r)$ when the coat is of size $r_0 = 10\;nm$ (left) and the change in membrane energy-per-area of coat monomers $\Delta f_c$ due to coupling to the composition field against the coat radius $r_0$ (right). In both cases we assume a typical intrinsic coat curvature with $\mathcal{R}_c = 50\;nm$, $\beta = 1.0\;nm^{-1}$ and $\gamma = 1.1\;nm^{-1}$.  In the underdamped regime the energy change $\Delta f_c$ can exhibit an initial decrease, which may drive coat assembly. In both figures (a) and (b) we use $\kappa = 20\;k_B T$.}
\end{figure}

Mechanosensitive membrane channels have been widely studied and reveal the interplay between the biological function of transmembrane proteins and the adjacent membrane structure and composition. Through conformational changes from a closed to an open state that allows the passage of solvent through the membrane, they can equilibrate an osmotic imbalance between the interior and exterior of cells \cite{Perozo2006, Booth2007, Haswell2011}. Although many examples of these channels are found in nature, the bacterial mechanosensitive channels of large (MscL) and small conductance (MscS) are prototypes of such proteins. Experimental studies have have shown that the channel opening probability is related to the membrane tension and the size of the open pore \cite{Sukharev1994, Perozo2006, Booth2007, Haswell2011, Kung2010, Chiang2004, Sukharev2002, Perozo2002}. One possibility is that the channel simply dilates open at high tension but the transition between the closed and open states might also involve, e.g.~a change in slope at the protein-membrane interface (here, $\delta$) \cite{Turner2004}. In a two-component membrane such a change in boundary conditions between the closed and open states couples to both the shape and asymmetry field in the nearby membrane and hence contributes to a change in the free energy of the channel-membrane system. Here, for simplicity, we consider the angle at the channel wall $\delta$ to be non-zero in a conical closed state and $\delta = 0$ in the open state. We explore the thermodynamic effect of this gating-by-tilt by comparing the deformation energy $\mathcal{F}$ of the membrane to the experimental estimate of the energy required to open the channels at zero tension, inferred by assuming purely dilational opening. Fig~4(a) shows that the even modest changes in the boundary angle at the face of the channel could give rise to a significant thermodynamic energy under gating-by-tilt. Moreover, a regime is identified in which the membrane can act to \textit{close}, rather than open, the channel, characterized by a negative total energy $\mathcal{F}$ relative to the open state, although a similar result was previously identified in a model that neglects spatial variation of coupling to curvature \cite{Wiggins2004, Wiggins2005}. Our results indicate that lipid composition variation, and its coupling to mean curvature, could play a role in regulating the function of mechanosensitive membrane channels.

Finally, the presence of a negative deformation energy in the underdamped regime motivated us to study the thermodynamics of protein coat formation on a membrane. Such coats are important in regulation, e.g. membrane trafficking using clathrin coats, or in infection, where viral coats assemble at the plasma membrane \cite{molecular-bio}. Fig~4(b) shows both the compositional field $\phi$ around a protein coat of size $r_0=10\;nm$ and the variation with $r_0$ of $\Delta f_c$, i.e.~the change in membrane energy due to coupling to $\phi$ only, scaled by the coat area. Thus, $\Delta f_c$ renormalizes the chemical potential for binding of early coat monomers to the membrane, and it is computed by adding both the contribution from the membrane inside (under) and outside the coat \cite{Note2}. Two striking features are observed in the underdamped regime. Firstly, this free-energy change can support an initial \textit{decrease} with coat size. In this case the deformation of the membrane (with its associated composition) is energetically {favorable}. This is true (even) for membranes that remain thermodynamically stable, i.e. in the absence of  bulk instability. This may represent a new mechanism for driving (controlling) coat formation in cells. This might be tested by tracking coat assembly at different $\alpha$ (tension), e.g. controlled by micropipette aspiration \cite{Evans1990}: We predict a dramatic increase in the rate of assembly near the critical tension $\sigma_c$. Secondly, the existence of a minimum in $\Delta f_c$ corresponds to a characteristic coat size with metastable character; we note that partially formed coats are often observed \cite{Heuser1980}.

\begin{acknowledgments}
We acknowledge the stimulating discussions with Dr P. Sens (Paris) and Profs M. Freissmuth and H. Sitte (Vienna) and funding from EPSRC under grant EP/E501311/1 (a Leadership Fellowship to MST). 
\end{acknowledgments}

\end{document}


\begin{flushleft}
{\textbf{\textsl{\huge Supporting Information: \\[1ex] \Large Membrane Composition Variation and Underdamped Mechanics near Transmembrane Proteins and Coats }}}\\[+2.5ex]

S. Alex Rautu$^{1}$, 
George Rowlands$^{1}$, 
Matthew S. Turner$^{1,2,\ast}$
\\ \vspace{5pt}
\it ${}^1$\,Department of Physics, University of Warwick, Coventry, CV4 7AL, UK
\\[-0.5ex]
\it ${}^2$\,Centre for Complexity Science, University of Warwick, Coventry, CV4 7AL, UK
\\[+0.0ex]
${}^\ast$\,e-mail: m.s.turner@warwick.ac.uk 
\end{flushleft}

\section{Euler-Lagrange Equations and Boundary Conditions}

The effective free-energy functional of a two-component lipid bilayer near a transmembrane inclusion that describes the mid-plane deviations from flatness, $u(\mathbf{r})$, and its corresponding local compositional asymmetry, $\phi(\mathbf{r})$, as described in the main text, is given by \begin{equation}\label{eqn:rF}
    \tilde{\mathcal{F}} \equiv \frac{\mathcal{F}}{\pi\kappa} = \frac{1}{2\pi}\int_{\mathcal{M}} \mathrm{d}^2\mathbf{r}\left[\,\alpha^2\,(\nabla u)^2 + (\nabla^2 u)^2 + \gamma^2\left(\,\beta^2\,\phi^2 + (\nabla \phi)^2 + 2\,\phi\,(\nabla^2 u)\,\right)\,\right]\!,
\end{equation} which is nondimensionalized by the membrane bending rigidity $\kappa$. Here, $\mathcal{M}$ is the region of integration, that is, a Cartesian plane without the cross-sectional area of the protein inclusion, $\mathbf{r}$ is a position vector within this plane, and the differential operators $\nabla$ and $\nabla^2$ are defined by the two-dimensional versions of the usual gradient and Laplacian operator, respectively. Also, the free parameters $\alpha$, $\beta$ and $\gamma$ have dimensions of inverse length and they represent the typical length scales within our model. 

In order to derive the differential equations that $u(\mathbf{r})$ and $\phi(\mathbf{r})$ must obey to minimize (\ref{eqn:rF}), the so-called Euler-Lagrange equations \cite{Riley2006}, we consider the one-parameter family of maps \begin{equation}\label{eqn:maps}u(\mathbf{r})\rightarrow u_\star(\mathbf{r}) \equiv u(\mathbf{r}) + \epsilon_u\,\delta u(\mathbf{r})\qquad\text{and}\qquad\phi(\mathbf{r})\rightarrow \phi_\star(\mathbf{r}) \equiv \phi(\mathbf{r}) + \epsilon_\phi\,\delta \phi(\mathbf{r}),\end{equation} where the test functions $\delta u(\mathbf{r})$ and $\delta \phi(\mathbf{r})$ are completely arbitrary within $\mathcal{M}$, but constrained at the boundaries of the domain $\partial\mathcal{M}$. Hence, a quantity $\tilde{\mathcal{F}}_\star(\epsilon_u, \epsilon_\phi)$ analogous to $\tilde{\mathcal{F}}$ of equation (\ref{eqn:rF}) can be constructed with these transformations \cite{Byron1992}. By setting $\epsilon_u=\epsilon_\phi = 0$ to be the point in the parameter space that labels the desired minimizing functions, then $\tilde{\mathcal{F}}_\star$ is required to be a minimum with respect to $\epsilon_u$ and $\epsilon_\phi$ at this point, namely \begin{equation}\label{eqn:limits}\lim_{\epsilon_u\rightarrow\,0} \frac{\tilde{\mathcal{F}}_\star (\epsilon_u, 0)-\tilde{\mathcal{F}}}{\epsilon_u} = \lim_{\epsilon_\phi\rightarrow\,0} \frac{\tilde{\mathcal{F}}_\star (0,\epsilon_\phi)-\tilde{\mathcal{F}}}{\epsilon_\phi}=0.\end{equation} 

The explicit form of $\tilde{\mathcal{F}}_\star$ can be found by applying (\ref{eqn:maps}) onto (\ref{eqn:rF}), which can be written to lowest order in $\epsilon_u$ and $\epsilon_\phi$ as follows:  
\begin{align}
\tilde{\mathcal{F}}_\star (\epsilon_u, \epsilon_\phi) = \tilde{\mathcal{F}} &+ \frac{\epsilon_u}{\pi}\int_{\mathcal{M}}\mathrm{d}^2\mathbf{r}\left(\,\nabla^2 u\,\nabla^2\delta u + \alpha^2\,\nabla u \cdot\nabla\delta u +\gamma^2\,\phi\,\nabla^2\delta u\,\right) +\mathcal{O}(\epsilon^2_u) \notag \\ 
&\qquad\qquad\qquad+ \frac{\epsilon_\phi}{\pi}\int_{\mathcal{M}}\mathrm{d}^2\mathbf{r}\,\,\gamma^2\left(\,\beta^2\phi\,\delta\phi + \nabla\phi\cdot\nabla\delta\phi + \delta\phi\,\nabla^2 u\,\right) +\mathcal{O}(\epsilon^2_\phi).\end{align} 

As a result, equation (\ref{eqn:limits}) yields that \begin{equation}\mathcal{I}_u\equiv\int_{\mathcal{M}}\mathrm{d}^2\mathbf{r}\left(\,\nabla^2 u\,\nabla^2\delta u + \alpha^2\,\nabla u \cdot\nabla\delta u +\gamma^2\,\phi\,\nabla^2\delta u\,\right) = 0\end{equation} and \begin{equation}\mathcal{I}_\phi\equiv\int_{\mathcal{M}}\mathrm{d}^2\mathbf{r}\left(\,\beta^2\phi\,\delta\phi + \nabla\phi\cdot\nabla\delta\phi + \delta\phi\,\nabla^2 u\,\right) = 0.\vspace{5pt}\end{equation} 

By employing the divergence theorem and the method of integration by parts \cite{Riley2006}, these integrals can be reduced to \begin{align}\label{eqn:EL1}\mathcal{I}_\phi = \oint_{\partial\mathcal{M}}\!\delta \phi(\mathbf{r})\,\frac{\partial \phi}{\partial n} - \int_{\mathcal{M}}\mathrm{d}^2\mathbf{r}\,\left((\nabla^2-\beta^2)\phi-\nabla^2 u\,\right)\delta \phi(\mathbf{r}) = 0 \end{align} and \begin{align}\label{eqn:EL2}\mathcal{I}_u 
&= \oint_{\partial\mathcal{M}}\!\delta u(\mathbf{r})\,\frac{\partial}{\partial n}\!\left(\alpha^2\,u - \gamma^2\,\phi-\nabla^2 u\right) + \oint_{\partial\mathcal{M}}\left(\nabla^2 u + \gamma^2\,\phi\,\right)\frac{\partial}{\partial n}\delta u(\mathbf{r}) \notag \\[2ex] 
& \hspace{6.5cm}+\int_{\mathcal{M}}\mathrm{d}^2\mathbf{r}\,\left(\nabla^2\,(\nabla^2-\alpha^2) u + \gamma^2\,\nabla^2\phi\,\right)\delta u(\mathbf{r}) = 0,
\end{align} where the solutions for both $u(\mathbf{r})$ and $\phi(\mathbf{r})$ are imposed to vanish in the far-field limit, and $\frac{\partial}{\partial n}\equiv\hat{\textbf{n}}\cdot\nabla$ denotes the normal derivative, with $\hat{\textbf{n}}$ as the unit vector normal to the boundary $\partial\mathcal{M}$ and pointing toward the center of the membrane inclusion.

To determine the Euler-Lagrange equations, the boundary terms in (\ref{eqn:EL1}) and (\ref{eqn:EL2}) are required to cancel out \cite{Byron1992}. In the latter case, this can be easily achieved by demanding that both $\delta u$ as well as the normal component of its gradient, $\frac{\partial}{\partial n}\delta u$, vanish everywhere on the boundary. However, from (\ref{eqn:maps}), this is equivalent to fixing the value of the field $u(\mathbf{r})$ and its normal derivative on $\partial\mathcal{M}$. This choice of boundary conditions is motivated by the strong coupling between the transmembrane domain of the inclusion and the hydrophobic core of the membrane. Similarly, the vanishing of the boundary term in (\ref{eqn:EL1}) can be attained by setting the normal derivative, $\frac{\partial}{\partial n}\phi$, to be zero everywhere on $\partial\mathcal{M}$ (Neumann condition), or by requiring that $\delta\phi = 0$, that is, $\phi$ is fixed on the boundary (Dirichlet condition). 

Hereinafter, we choose the following boundary conditions: \begin{equation}\label{eqn:BCs}
  u(\mathbf{r})\,\Big|_{\partial\mathcal{M}} = \,\mathcal{U},\qquad\text{and}\qquad
  \mathbf{\hat{n}}\cdot\nabla u\,\Big|_{\partial\mathcal{M}} = \,\mathcal{U}',
\end{equation} where $\partial\mathcal{M}$ is assumed to be a simple connected curve, and the functions $\,\mathcal{U}$ and $\,\mathcal{U}'$ describe the height and the contact angle at which the mid-plane of the bilayer meets the membrane inclusion, respectively. Regardless of the choice of boundary condition on $\phi(\mathbf{r})$, and since the test functions $\delta u(\mathbf{r})$ and $\delta \phi(\mathbf{r})$ are, by definition, completely arbitrary in the bulk, the equations (\ref{eqn:EL1}) and (\ref{eqn:EL2}) can be simplified to \begin{equation}\label{eqn:EL}\nabla^2 u=(\nabla^2 - \beta^2)\phi\quad\text{and}\quad\nabla^2(\nabla^2 - \alpha^2)\,u\,+\,\gamma^2\,\nabla^2\phi = 0,\end{equation} respectively, where (\ref{eqn:BCs}) is used to remove the boundary terms in (\ref{eqn:EL2}). These are the Euler-Lagrange equations that  $u(\mathbf{r})$ and $\phi(\mathbf{r})$ must satisfy in the lowest energy state of the system.

\section{Derivation of the Equilibrium Solutions}

At thermodynamic equilibrium, the effective free-energy $\mathcal{F}$ must reach a minimum with respect to $u(\mathbf{r})$, the deviation from flatness of the bilayer mid-plane, and the order parameter $\phi(\mathbf{r})$, the local compositional asymmetry. Therefore, using the Euler-Lagrange equations as determined in the previous section, they can be combined to give a single differential equation for $\phi$, as follows:
\begin{equation}
	\left[\nabla^4 - (\alpha^2 + \beta^2 -\gamma^2)\nabla^2 + \alpha^2\beta^2\right]\phi = 0,
\end{equation} which can be written in the more convenient form 
\begin{equation}
	\label{eqn:phi-diff}
	\left(\nabla^2 - k_+^2\right)\left(\nabla^2 - k_-^2\right)\phi=0,
\end{equation} 
by setting $\alpha^2 + \beta^2 -\gamma^2 = k^2_+ +k^2_-$ and $\alpha^2\beta^2 = k^2_+\,k^2_-$, where the expressions for $k^2_\pm$ can be found, in terms of $\alpha$, $\beta$ and $\gamma$, to be 
\begin{equation}
	\label{eqn:kSq-eqn}
	k^2_\pm = \frac{1}{2}\left[\,(\alpha^2 + \beta^2 -\gamma^2)\,\pm\,\sqrt{(\alpha^2 + \beta^2 -\gamma^2)^2 - 4\alpha^2\beta^2\,}\,\right]\!.
\end{equation} 

By inspection, it can be seen that $k^2_\pm>0$ if and only if $\gamma\in\left(0,\,|\alpha-\beta|\,\right]$ and also that $k^2_\pm<0$ if and only if $\gamma\in\left[\,|\alpha+\beta|, \infty\right)$. Moreover, there are no opposite sign solutions for $k^2_\pm$ and for $\gamma\in\left(\,|\alpha-\beta|,|\alpha+\beta|\,\right)$, the solutions are found to be complex. Consider, for now, the case when $0<\gamma<|\alpha-\beta|$, so that $k^2_\pm$ are strictly positive with $k_-\neq k_+$. By defining $\phi_\pm$ such that $\phi_+ = (\nabla^2 - k_-^2)\phi$ and $\phi_- = (\nabla^2 - k_+^2)\phi$, then we have that 
\begin{equation}
	\label{eqn:phi-pm1}
	\phi = \frac{\,\phi_+ - \phi_-\,}{\,k_+^2-k_-^2\,}.
\end{equation} 

Consequently, by linearity, the solution of equation (\ref{eqn:phi-diff}) can be found by solving separately two differential equations of the form 
\begin{equation}
	\label{eqn:phi-pm2}
	\left(\nabla^2 - k_\pm^2\right)\phi_\pm=0.
\end{equation} 
This equation is the two-dimensional version of the Helmholtz equation \cite{Riley2006}, which can be written in polar coordinates as 
\begin{equation}
	\label{eqn:r-phi-diff}
	\frac{\partial^2\phi_\pm}{\partial\, r^2} + \frac{1}{r}\frac{\partial\phi_\pm}{\partial r} + \frac{1}{r^2}\frac{\partial^2\phi_\pm}{\partial\,\theta^2} - k^2_\pm\,\phi_\pm  = 0.
\end{equation} 

Equation (\ref{eqn:r-phi-diff}) can be solved by separation of variables, that is, $\phi_\pm(r, \theta) = R^\pm(r)\,\Theta^\pm(\theta)$, which gives by substitution the following differential equations: 
\begin{equation}
	\frac{\partial^2\Theta^\pm}{\partial\,\theta^2}+n^2\,\Theta^\pm = 0\quad\text{and}\quad r^2\,\frac{\partial^2 R^\pm}{\partial\,r^2}+r\,\frac{\partial R^\pm}{\partial r}-\left(k^2_\pm\,r^2 + n^2\right)R^\pm = 0,
\end{equation} 
where $n\in\mathbb{Z}$ (integer numbers). The family of  solutions for the first equation is of the form 
\begin{equation}
	\label{eqn:theta-sol}
	\Theta^\pm_n(\theta)=\mathcal{S}_n \sin{(n\theta)} + \mathcal{C}_n \cos{(n\theta)} \equiv \Theta_n(\theta),
\end{equation} 
where $\mathcal{S}_n$ and $\,\mathcal{C}_n\,$ are arbitrary real constants. The second equation is a familiar differential equation that admits as solutions the modified Bessel functions \cite{Abramowitz1965}. Thus, the general solution of $\phi_\pm$ is given by 
\begin{equation}
	\label{eqn:phi-sol-g}
	\phi_\pm(r, \theta)=\sum_{n\,\in\,\mathbb{Z}}\left[A_n^\pm\,K_n(k_\pm\,r)+B_n^\pm\,I_n(k_\pm\,r)\right]\Theta_n(\theta),
\end{equation} 
where $A^\pm_n$ and $B^\pm_n$ are constants, and the functions $I_n$ and $K_n$ are the $n$-th order modified Bessel functions of the first and second kind, respectively \cite{Abramowitz1965}. Since $\displaystyle\lim_{x\to\infty}I_n (x) = \infty$ for any $n$ \cite{Abramowitz1965, Gradshteyn2000}, and only solutions that vanish in the far-field limit are of interest, then we require that $B^\pm_n =0$. Consequently, this yields a solution for equation (\ref{eqn:phi-diff}) as follows: 
\begin{equation}
	\phi(r, \theta)=\sum_{n\,\in\,\mathbb{Z}}\,\frac{A^{+}_{n}\,K_n(k_+\,r) - A^{-}_{n}\,K_n(k_-\,r)}{k^2_+-k^2_-}\,\Theta_n(\theta).
\end{equation}

As a result, the conformation of the mid-plane of the bilayer, $u(r,\theta)$, can be determined from the first Euler-Lagrange equation in (\ref{eqn:EL}), namely
\begin{equation}
	\nabla^2u(r,\theta)=\left(\nabla^2-\beta^2\right)\phi(r,\theta) = \frac{\left(k^2_{+}-\beta^2\right)\phi_{+}(r,\theta)-\left(k^2_{-}-\beta^2\right)\phi_{-}(r,\theta)}{k^2_+-k^2_-},
\end{equation} 
where the last equality is obtained using (\ref{eqn:phi-pm1}) and (\ref{eqn:phi-pm2}). This leads to the following inhomogeneous differential equation: 
\begin{equation}
	\label{eqn:lap-u-diff}
	\nabla^2\,u(r,\,\theta)=\sum_{n\,\in\,\mathbb{Z}}\left[k^2_{+}U^{+}_{n}\,K_n(k_+\,r)+k^2_{-}U^{-}_{n}\,K_n(k_-\,r)\right]\Theta_n(\theta),
\end{equation} 
where the constants $U^{\pm}_{n}$ are defined by 
\begin{equation}
	\label{eqn:def-par-U}
	U^\pm_n=\frac{\pm\,A^\pm_n}{\,k^2_+-k^2_-\,}\left(\,1-\frac{\beta^2}{k^2_\pm}\,\right)\!.
\end{equation}

The general solution of (\ref{eqn:lap-u-diff}) is found by employing the ansatz 
\begin{equation}
	u(r,\theta) = u_{+}(r,\theta)+u_{-}(r,\theta)+u_{h}(r,\theta),
\end{equation} 
where $u_h$ is the solution that solves the two-dimensional Laplace equation \cite{Riley2006}, i.e. $\nabla^2u_h(r,\theta)=0$, and the functions $u_\pm(r,\theta)$ are given by 
\begin{equation}
	\label{eqn:inh-sol-diff}
	\nabla^2u_{\pm}(r,\theta) = k^2_\pm\sum_{n\,\in\,\mathbb{Z}}U^{\pm}_{n}\,K_n(k_\pm\,r)\,\Theta_n(\theta).
\end{equation} 

The inhomogeneous solution of (\ref{eqn:inh-sol-diff}) is found to be 
\begin{equation}
	u_{\pm}(r,\theta) = \sum_{n\,\in\,\mathbb{Z}}U^{\pm}_{n}\,K_n(k_\pm\,r)\,\Theta_n(\theta),
\end{equation} 
which, analogous to the previous result in equation (\ref{eqn:phi-pm2}), satisfies $(\nabla^2-k^2_\pm)\,u_{\pm} = 0$. On the other hand, the homogeneous solution $u_h$ can be written as 
\begin{equation}
	u_h(r,\theta)=W'\log(r) + \sum_{n\,\in\,\mathbb{Z}}W_n\,r^{-n}\,\Theta_n(\theta),
\end{equation} 
where $W'$ and $W_n$ are arbitrary constants. As before, we need to rule out the terms in $u(r,\theta)$ that diverge at infinity, that is, $W_n = 0$ for $n<0$. The logarithmic solution corresponds physically to membrane proteins that exert a normal force on the membrane. Since the membrane inclusions are assumed to be free, this solution is also discarded, which yields that 
\begin{equation}
	\label{eqn:u}
	u(r,\theta)=\sum^{\infty}_{n=0}\mathcal{V}^+_n(\theta)\,K_n(k_+ r) + \mathcal{V}^-_n(\theta)\,K_n(k_- r)+\mathcal{W}_n(\theta)\,r^{-n},
\end{equation} 
where the identity $K_n(x) = K_{-n}(x)$ for any $x$ \cite{Abramowitz1965, Gradshteyn2000}, is used, and the circular functions $\mathcal{V}^\pm_n(\theta)$ and $\mathcal{W}_n(\theta)$ are defined by 
\begin{equation}
	\label{eqn:Vpm}
	\mathcal{V}^\pm_n(\theta) = \left\{\begin{array}{l l}
    U^\pm_{0}\,\Theta_{0}(\theta), & \text{if \,$n=0$} \\[1ex]
    U^\pm_{n}\,\Theta_{n}(\theta) + U^\pm_{-n}\Theta_{-n}(\theta), & \text{if \,$n>0$}
	\end{array} \right.
\end{equation} and \begin{equation}\mathcal{W}_n(\theta) = W_{n}\,\Theta_{n}(\theta),\quad\text{with $n\geq0$}.\vspace{1ex}\end{equation} 

Thus, the local compositional asymmetry $\phi(r,\theta)$ can be expressed, as well, in terms of (\ref{eqn:Vpm}), which becomes
\begin{equation}
	\label{eqn:phi}
	\phi(r,\theta)=\sum^{\infty}_{n=0}f_+\mathcal{V}^+_n(\theta)\,K_n(k_+ r) + f_-\mathcal{V}^-_n(\theta)\,K_n(k_- r),
\end{equation} 
where the coefficients $f_\pm$ are given by 
\begin{equation}
	f_\pm = \frac{k^2_\pm}{k^2_\pm-\beta^2}.
\end{equation}

\section{Neumann Boundary Condition}

By assuming cylindrical symmetry for the membrane inclusion (say, of radius $r_0$), and choosing the Neumann condition for $\phi(\mathbf{r})$ on $\partial\mathcal{M}$ (later, the Dirichlet condition will be examined as well), the unknown functions $\mathcal{V}^\pm_n$ and $\mathcal{W}_n$ can be obtained from the boundary conditions, which can be written as 
\begin{equation}
\label{eqn:BC-phi}\mathbf{\hat{n}}\cdot \nabla \phi(r_0, \theta) = 0,
\end{equation} 
\begin{equation}
\label{eqn:BC-u}u(r_0, \theta)= \,\mathcal{U}(\theta)\equiv \sum_{n=0}^{\infty} z_n \cos(n\theta - \xi_n),\end{equation} and 
\begin{equation}\label{eqn:BC-u'}\mathbf{\hat{n}}\cdot \nabla u(r_0, \theta) = \,\mathcal{U}'(\theta)\equiv \sum_{n=0}^{\infty} \delta_n \cos(n\theta - \chi_n),
\end{equation} where $\mathbf{\hat{n}}$ is the inward unit vector normal to the boundary and $\xi_0 = \chi_0 = 0$. The functions $\,\mathcal{U}(\theta)$ and $\,\mathcal{U}'(\theta)$ are the height and the contact angle at which the membrane meets the protein inclusion, respectively. Moreover, they are given as a Fourier series \cite{Riley2006}, so that $\mathcal{V}^\pm_n(\theta)$ and $\mathcal{W}_n(\theta)$ can be evaluated at each order in the series (\ref{eqn:u}) and (\ref{eqn:phi}). Thus, these conditions can be re-written in a matrix form as follows:
\begin{equation}
	\label{eqn:BC-matrix}
	\begin{pmatrix}
  	\,K_n(k_+r_0)\, 			& \,K_n(k_-r_0)\, 			& \,1\, \\[1ex]
  	\,\mathcal{L}^+_n K_n(k_+r_0)\, 	& \,\mathcal{L}^-_n K_n(k_-r_0)\, 	& \,n\, \\[1ex]
  	\,f_+\mathcal{L}^+_n K_n(k_+r_0)\, 	& \,f_-\mathcal{L}^-_n K_n(k_-r_0)\, 	& \,0\,
 	\end{pmatrix}\begin{pmatrix}\mathcal{V}^+_n(\theta) \\[1ex] \mathcal{V}^-_n(\theta) \\[1ex] \mathcal{W}(\theta)\,r_0^{-n}\end{pmatrix} 
 	= \begin{pmatrix}z_n\cos(n\theta-\xi_n) \\[1ex] r_0\,\delta_n\cos(n\theta-\chi_n) \\[1ex] 0 \end{pmatrix}\!,
\end{equation} with $n\geq0\,$ and 
\begin{equation}
	\label{eqn:Lpm}
	\mathcal{L}^\pm_n = -\frac{k_\pm r_0\,K'_n(k_\pm r_0)}{K_n(k_\pm r_0)}  = n + \frac{k_\pm r_0\,K_{n-1}(k_\pm r_0)}{K_n(k_\pm r_0)},
\end{equation} where the last step is obtained by using a recurrence relation of the modified Bessel functions \cite{Abramowitz1965, Gradshteyn2000}:
\begin{equation}
	K'_n(x) = -K_{n-1}(x) - \frac{n}{x}\,K_n(x),
\end{equation} for any $x$, where the prime symbol denotes here a derivative with respect to the argument of the function. 

In order to determine $\mathcal{V}^\pm_n(\theta)$ and $\mathcal{W}_n(\theta)$, the matrix in equation (\ref{eqn:BC-matrix}), say $M$, must be invertible, namely its determinant is required to be non-zero \cite{Riley2006}. This implies that 
\begin{equation}
	\mathcal{N}_n \equiv f_-\mathcal{L}^-_n(n-\mathcal{L}^+_n)-f_+\mathcal{L}^+_n(n-\mathcal{L}^-_n) \neq 0,\vspace{-1ex}
\end{equation} and 
\begin{equation}\vspace{-2ex}
\label{eqn:inv-matrix-eq}
\frac{1}{\mathcal{N}_n}\begin{pmatrix}
 	\,\frac{n f_- \mathcal{L}^-_n}{K_n\left(k_+ r_0\right)}\, 	& -\frac{f_- \mathcal{L}^-_n}{K_n\left(k_+ r_0\right)} 	&	\frac{\mathcal{L}^-_n-n}{K_n\left(k_+ r_0\right)}\, \\[1ex]
 	\,-\frac{n f_+ \mathcal{L}^+_n}{K_n\left(k_- r_0\right)}\, 	& \frac{f_+ \mathcal{L}^+_n}{K_n\left(k_- r_0\right)} 	& \frac{n-\mathcal{L}^+_n}{K_n\left(k_- r_0\right)}\, \\[1ex]
 	\,\mathcal{L}^+_n \mathcal{L}^-_n\left(f_+-f_-\right)\,			& f_- \mathcal{L}^-_n-f_+ \mathcal{L}^+_n 							& \mathcal{L}^+_n-\mathcal{L}^-_n\,
\end{pmatrix}\begin{pmatrix}z_n\cos(n\theta-\xi_n) \\[1ex] r_0\,\delta_n\cos(n\theta-\chi_n) \\[1ex] 0 \end{pmatrix}
=\begin{pmatrix}\mathcal{V}^+_n(\theta) \\[1ex] \mathcal{V}^-_n(\theta) \\[1ex] \mathcal{W}(\theta)\,r_0^{-n}\end{pmatrix}\!,\vspace{+2ex}
\end{equation} which is obtained by multiplying (\ref{eqn:BC-matrix}) by the inverse of the matrix $M$. Hence, we have that
\begin{equation}
\label{eqn:Vn}
\mathcal{V}^{\pm}_{n}(\theta) = \pm\frac{f_\mp \mathcal{L}^{\mp}_{n}}{\mathcal{N}_{n}\,K_{n}(k_\pm r_0)}\left[n z_n \cos(n\theta - \xi_n)- r_0 \delta_n \cos(n\theta - \chi_n)\right]
\end{equation} and 
\begin{equation}
\label{eqn:Wn}
\mathcal{W}_n(\theta) = \frac{1}{r^{-n}_0\mathcal{N}_{n}}\left[\mathcal{L}^{+}_{n}\mathcal{L}^{-}_{n}\left(f_{+}-f_{-}\right) z_n \cos(n\theta - \xi_n)- \left(f_{+} \mathcal{L}^{+}_{n} - f_{-} \mathcal{L}^{-}_{n}\right)r_0 \delta_n \cos(n\theta - \chi_n)\right]\!.
\end{equation}

Therefore, the membrane profile, $u(r,\theta)$, and its compositional asymmetry, $\phi(r,\theta)$, are completely determined by the boundary conditions (\ref{eqn:BC-phi}\,--\,\ref{eqn:BC-u'}). However, the form of $\,\mathcal{U}(\theta)$ and $\,\mathcal{U}'(\theta)$ are not arbitrary, as the height $z_0$ is set by the minimization of the effective free-energy (\ref{eqn:rF}), which, in fact, gives the condition for the balance of vertical forces. Hence, using the explicit form of (\ref{eqn:u}), (\ref{eqn:Vn}) and (\ref{eqn:Wn}), and the requirement that the membrane profile, $u(r,\theta)$, vanishes in the far-field limit, then we obtain that
\begin{equation}
	\label{eqn:z0}
	z_0 = r_0\,\delta_0\,\frac{f_{+}\mathcal{L}^{+}_{0}-f_{-}\mathcal{L}^{-}_{0}}{\mathcal{L}^{+}_{0}\mathcal{L}^{-}_{0}\left(f_{+}-f_{-}\right)}.
\end{equation} 

Another equilibrium condition that the system needs to fulfill is the balance of torques on the rigid inclusion, which is explored in the next section.

\section{Membrane Deformation Energy and Torque Balance}

By employing the divergence theorem and the vector identity $\nabla\mathcal{A}\cdot\nabla\mathcal{B} = \nabla\cdot\left(\mathcal{A}\,\nabla\mathcal{B}\right) - \mathcal{A}\,\nabla^2\mathcal{B}$, with $\mathcal{A}$ and $\mathcal{B}$ as two scalar fields, then
\begin{equation}
	\int_\mathcal{M}\!\mathrm{d}^2\mathbf{r}\,\,(\nabla u)^2 = \oint_{\partial\mathcal{M}}u\,\frac{\partial u}{\partial n} - \int_\mathcal{M}\!\mathrm{d}^2\mathbf{r}\,\,u\left(\nabla^2 u\right)\!,
\end{equation} where, as before, $\frac{\partial}{\partial n}\equiv\hat{\textbf{n}}\cdot\nabla$ denotes the normal derivative. Similarly, an identical relationship can be obtained for the scalar field $\phi(\mathbf{r})$. Thus, the free-energy given by (\ref{eqn:rF}) can be expressed as
\begin{equation}
	\tilde{\mathcal{F}} = \frac{1}{2\pi}\oint_{\partial\mathcal{M}}\left(\alpha^2 u\,\frac{\partial u}{\partial n} + \gamma^2 \phi\,\frac{\partial \phi}{\partial n}\,\right) + \frac{1}{2\pi}\int_{\mathcal{M}} \mathrm{d}^2\mathbf{r}\,\,(\nabla^2 u)\left[(\nabla^2 - \alpha^2)\,u\,+\,\gamma^2\phi\,\right]\!,
\end{equation} where the last term is reduced to this form with the help of the first Euler-Lagrange equation in (\ref{eqn:EL}). By denoting the term in square brackets by $w$ and using the identity $\nabla(w\nabla u - u\nabla w) = w\nabla^2 u -u \nabla^2 w$, this yields that
\begin{equation}
	\tilde{\mathcal{F}} = \frac{1}{2\pi}\oint_{\partial\mathcal{M}}\left(\alpha^2 u\,\frac{\partial u}{\partial n} + \gamma^2 \phi\,\frac{\partial \phi}{\partial n} + w\,\frac{\partial u}{\partial n} - u\,\frac{\partial w}{\partial n}\,\right)\!,
\end{equation}
where the second Euler-Lagrange equation, i.e.\ $\nabla^2 w = 0$, has been used to remove the bulk terms and rewrite the free-energy as a line integral over the boundary $\partial\mathcal{M}$. The equilibrium solutions for $u$ and $\phi$, as derived in the previous section, are given by
\begin{equation}
	\label{eqn:u-phi-ansatz}
	\phi(r,\theta)=f_+u_+(r,\theta) + f_-u_-(r,\theta)\quad\text{and}\quad u(r,\theta)=u_+(r,\theta) + u_-(r,\theta) + u_h(r,\theta),
\end{equation} in which the coefficients $f_\pm = k^2_\pm\,/\,(k^2_\pm-\beta^2)$ and the functions $u_h$ and $u_\pm$ satisfies 
\begin{equation}
	\label{eqn:u-phi-conditions}
	\nabla^2u_h=0\quad\text{and}\quad(\nabla^2-k^2_\pm)\,u_{\pm} = 0,
\end{equation} respectively. Using (\ref{eqn:u-phi-ansatz}) and (\ref{eqn:u-phi-conditions}), $w$ is found to be 
\begin{equation}
	w(r,\theta) = -\alpha^2 u_h(r,\theta) = -\alpha^2\sum^{\infty}_{n=0} \mathcal{W}_n(\theta)\,r^{-n},
\end{equation} where the last step is obtained by employing the explicit form of $u_h$. Hence, this yields that
\begin{equation}
	\label{eqn:noFBCs0}
	\tilde{\mathcal{F}} = \frac{\,\alpha^2}{2\pi}\oint_{\partial\mathcal{M}} \left(u\,\frac{\partial u}{\partial n} + u\,\frac{\partial u_h}{\partial n} - u_h\,\frac{\partial u}{\partial n}\right) + \frac{\gamma^2}{2\pi}\oint_{\partial\mathcal{M}} \phi\,\,\frac{\partial \phi}{\partial n},
\end{equation} which can be further simplify, by using the boundary conditions (\ref{eqn:BC-phi}\,--\,\ref{eqn:BC-u'}), as follows:
\begin{equation}
	\label{eqn:FBCs0}
	\tilde{\mathcal{F}} = \frac{\,\alpha^2}{2\pi}\int\limits^{2\pi}_{0}\mathrm{d}\theta\left[\,r_0\,\mathcal{U}'(\theta)\,\mathcal{U}(\theta) + \sum^{\infty}_{n=0}\mathcal{W}_n(\theta)\,r^{-n}_0\Big(n\,\mathcal{U}(\theta)-r_0\,\mathcal{U}'(\theta)\Big)\right]\!.
\end{equation}

Since the Euler-Lagrange equations (\ref{eqn:EL}) only express the balance of normal forces on any infinitesimal patch of the membrane, then the minimized energy, given by (\ref{eqn:FBCs0}), must also satisfy the equilibrium condition of zero net torque on the rigid inclusion. This condition implies that (\ref{eqn:FBCs0}) needs to be invariant under the transformations \begin{equation}\label{eqn:balance}\mathcal{U}(\theta)\mapsto\mathcal{U}(\theta) - r_0\psi\cos(\theta-\varepsilon)\quad\text{and}\quad\mathcal{U}'(\theta)\mapsto\mathcal{U}'(\theta) + \psi\cos(\theta-\varepsilon),\end{equation} where $\psi$ is a small tilt angle about the vertical axis $\mathbf{\hat{z}}$ (see Figure 1 in the main text), and the azimuthal angle $\varepsilon$ specifies the orientation of this tilt. The small angle approximation is used here to neglect the deviations from the circular shape of the projected area that is spanned by the membrane inclusion. Under such transformations the membrane profile, $u(\mathbf{r})$, and its local compositional asymmetry, $\phi(\mathbf{r})$, are affected as well. Since the terms in (\ref{eqn:balance}) couple only to the first Fourier modes of (\ref{eqn:BC-u}) and (\ref{eqn:BC-u'}), then the contributions to $u(\mathbf{r})$ and $\phi(\mathbf{r})$ due to torque balance (denoted by $u_\text{tilt}(\mathbf{r})$ and $\phi_\text{tilt}(\mathbf{r})$, respectively), can be determined by applying (\ref{eqn:balance}) to the matrix equation (\ref{eqn:inv-matrix-eq}), that is,
\begin{equation}
	\label{eqn:inv-matrix-tilt}
	\begin{pmatrix}\mathcal{V}^+_\text{tilt}(\theta) \\[1ex] \mathcal{V}^-_\text{tilt}(\theta) \\[1ex] \mathcal{W}_\text{tilt}(\theta)\,r_0^{-1}\end{pmatrix} = 
	\frac{1}{\mathcal{N}_1}\begin{pmatrix}
 	\,\frac{f_- \mathcal{L}^-_1}{K_1\left(k_+ r_0\right)}\, 	& -\frac{f_- \mathcal{L}^-_1}{K_1\left(k_+ r_0\right)} 	&	\frac{\mathcal{L}^-_1-1}{K_1\left(k_+ r_0\right)}\, \\[1ex]
 	\,-\frac{f_+ \mathcal{L}^+_1}{K_1\left(k_- r_0\right)}\, 	& \frac{f_+ \mathcal{L}^+_1}{K_1\left(k_- r_0\right)} 	& \frac{1-\mathcal{L}^+_1}{K_1\left(k_- r_0\right)}\, \\[1ex]
 	\,\mathcal{L}^+_1 \mathcal{L}^-_1\left(f_+-f_-\right)\,			& f_- \mathcal{L}^-_1-f_+ \mathcal{L}^+_1 							& \mathcal{L}^+_1-\mathcal{L}^-_1\,
\end{pmatrix}\begin{pmatrix}-r_0\psi\cos(\theta-\varepsilon) \\[1ex] r_0\psi\cos(\theta-\varepsilon) \\[1ex] 0 \end{pmatrix}\!,
\end{equation} where, as analogous to previous results in (\ref{eqn:u}) and (\ref{eqn:phi}), the azimuthal functions $\mathcal{V}^\pm_\text{tilt}(\theta)$ and $\mathcal{W}_\text{tilt}(\theta)$ prescribe the analytic form of
\begin{equation}
	\label{eqn:u-tilt}
	u_\text{tilt}(r,\theta)=\mathcal{V}^+_\text{tilt}(\theta)\,K_1(k_+ r) + \mathcal{V}^-_\text{tilt}(\theta)\,K_1(k_- r)+\mathcal{W}_\text{tilt}(\theta)\,r^{-1}
\end{equation} and \begin{equation}
	\label{eqn:phi-tilt}
	\phi_\text{tilt}(r,\theta)=f_+\mathcal{V}^+_\text{tilt}(\theta)\,K_1(k_+ r) + f_-\mathcal{V}^-_\text{tilt}(\theta)\,K_1(k_- r).
\end{equation} 

By applying the tilt transformations (\ref{eqn:balance}) and also $\mathcal{W}(\theta)\mapsto\mathcal{W}(\theta) + \mathcal{W}_\text{tilt}(\theta)$ onto (\ref{eqn:FBCs0}), and subsequently using the orthogonality relationships of the Fourier modes, the membrane deformation energy, in terms of the series (\ref{eqn:BC-u}) and (\ref{eqn:BC-u'}), becomes
\begin{equation}
\tilde{\mathcal{F}} = \sum^{\infty}_{n=0} \tilde{\mathcal{F}}_n + \tilde{\mathcal{F}}_\text{tilt},
\end{equation} where $\tilde{\mathcal{F}}_n$ is defined by 
\begin{equation}
	\label{eqn:rF-n}
	\tilde{\mathcal{F}}_n = \frac{\alpha^2}{\mathcal{N}_n\,\vartheta_n} \left[ r_0 \delta_n \left(f_{+} \mathcal{L}^{+}_{n} - f_{-} \mathcal{L}^{-}_{n}\right)\left(r_0 \delta_n -2 n z_n \cos(\xi_n-\chi_n)\right) +  n\,\mathcal{L}^{+}_{n}\mathcal{L}^{-}_{n}\left(f_{+}-f_{-}\right) z^2_n\right]\!,
\end{equation} with $\vartheta_0 = 1$ and $\vartheta_n = 2$ for any $n>0$. This gives the energy associated to each Fourier-Bessel mode, whereas the term given by $\tilde{\mathcal{F}}_\text{tilt}$ represents the energy contribution that accounts for the balance of torques, namely
\begin{equation}
	\label{eqn:rF-tilt}
	\tilde{\mathcal{F}}_\text{tilt} = \frac{\alpha^2}{2\,\mathcal{N}_1}\,\left[f_-\mathcal{L}^-_1(3+\mathcal{L}^+_1) - f_+\mathcal{L}^+_1(3+\mathcal{L}^-_1)\right]\!\left(z^2_\text{tilt} - 2\,z_\text{tilt}\,r_0 \delta_\text{tilt} \cos(\xi_1-\chi_1)+r^2_0\delta^2_\text{tilt}\right)\!,
\end{equation} which is obtained by minimizing the total energy with respect to $\psi$ and $\varepsilon$, where the minimum values obey the equations
\begin{equation}
	\label{eqn:minF-eqns}
	r_0\psi\cos(\varepsilon) = z_\text{tilt}\cos\xi_1 + r_0\delta_\text{tilt}\cos\chi_1\quad\text{and}\quad r_0\psi\sin(\varepsilon) = z_\text{tilt}\sin\xi_1 + r_0\delta_\text{tilt}\sin\chi_1.
\end{equation} The coefficients $\delta_\text{tilt}$ and $z_\text{tilt}$ are defined by
\begin{equation}
	\delta_\text{tilt} = \delta_1\,\frac{2\left(f_+\mathcal{L}^+_1 - f_-\mathcal{L}^-_1\right)}{f_+\mathcal{L}^+_1\left(3+\mathcal{L}^-_1\right) - f_-\mathcal{L}^-_1\left(3+\mathcal{L}^+_1\right)},
\end{equation} and 
\begin{equation}
	z_\text{tilt} = z_1\,\frac{f_+\mathcal{L}^+_1\left(1+\mathcal{L}^-_1\right) - f_-\mathcal{L}^-_1\left(1+\mathcal{L}^+_1\right)}{f_+\mathcal{L}^+_1\left(3+\mathcal{L}^-_1\right) - f_-\mathcal{L}^-_1\left(3+\mathcal{L}^+_1\right)},
\end{equation} respectively. Hence, using  (\ref{eqn:inv-matrix-tilt}) and (\ref{eqn:minF-eqns}), the azimuthal functions $\mathcal{V}^\pm_\text{tilt}(\theta)$ and $\mathcal{W}_\text{tilt}(\theta)$ are found to be
\begin{equation}
	\mathcal{V}^\pm_\text{tilt}(\theta) = \mp\frac{2\,f_\mp \mathcal{L}^\mp_1}{\mathcal{N}_1\,K_1\left(k_\pm r_0\right)}
	\big[z_\text{tilt} \cos(\theta - \xi _1) - r_0 \delta_\text{tilt}\cos(\theta - \chi _1)\big],
\end{equation} and 
\begin{equation}
	\mathcal{W}_\text{tilt}(\theta) = \frac{f_-\mathcal{L}^-_1\left(1+\mathcal{L}^+_1\right) - f_+\mathcal{L}^+_1\left(1+\mathcal{L}^-_1\right)}{r^{-1}_0\mathcal{N}_1}
	\big[z_\text{tilt} \cos(\theta - \xi _1) - r_0 \delta_\text{tilt}\cos(\theta - \chi _1)\big].
\end{equation}

As a result, by including the condition of zero net torque on the inclusion, the exact solutions of the membrane profile, $u(r, \theta)$, and its corresponding compositional asymmetry, $\phi(r, \theta)$, are given by 
\begin{equation}
	\label{eqn:final-u}
	u(r,\theta)=u_\text{tilt}(r,\theta) + \sum^{\infty}_{n=0}\mathcal{V}^+_n(\theta)\,K_n(k_+ r) + \mathcal{V}^-_n(\theta)\,K_n(k_- r)+\mathcal{W}_n(\theta)\,r^{-n},
\end{equation} and 
\begin{equation}
	\label{eqn:final-phi}
	\phi(r,\theta)=\phi_\text{tilt}(r,\theta) + \sum^{\infty}_{n=0}f_+\mathcal{V}^+_n(\theta)\,K_n(k_+ r) + f_-\mathcal{V}^-_n(\theta)\,K_n(k_- r), 
\end{equation} respectively.

\section{Curvature-induced Instabilities in Membranes}

Although the results of the previous section are derived only for $k^2_\pm > 0$ and $k_+ \neq k_-$, the solutions given by (\ref{eqn:final-u}) and (\ref{eqn:final-phi}) are not strictly restricted to this parameter space, but it can be analytically continued to a region where the values of $k^2_\pm$ are complex, namely $\gamma_d\leq\gamma<\gamma_c$, where $\gamma_d = |\alpha-\beta|$ and $\gamma_c = \alpha+\beta$. This can be easily seen by re-arranging equation (\ref{eqn:kSq-eqn}), which yields that
\begin{equation}
	\label{eqn:skpm}
	k_\pm = \frac{1}{2}\left[\sqrt{(\alpha+\beta)^2 - \gamma^2} \pm \sqrt{(\alpha-\beta)^2 - \gamma^2} \,\right]
\end{equation} are complex conjugates of each other within this interval. Furthermore, this guarantees that (\ref{eqn:final-u}) and (\ref{eqn:final-phi}) are real-valued solutions in this parameter region. However, this is no longer valid for a coupling term $\gamma>\gamma_c$, as the equation (\ref{eqn:phi-diff}) admits a completely different set of linearly independent solutions. 

The domain given by $\gamma>\gamma_c$ corresponds to a Leibler-type instability in membranes \cite{leibler1, leibler2}. This can be shown by analyzing the stability of the bulk of membranes as described by (\ref{eqn:rF}) far from any inclusions. Thus, by employing a Fourier transform of the total free-energy (\ref{eqn:rF}), say $\tilde{\mathcal{F}}_q$, then we have that
\begin{equation}
	\label{eqn:ftF}
	\tilde{\mathcal{F}}_q = \frac{1}{2\pi}\int\mathrm{d}^2\mathbf{q}\left[\left(q^4 + \alpha^2q^2\right) \left|u_q + \frac{\gamma^2}{\alpha^2+q^2}\,\phi_q\right|^2 + \gamma^2\left(\beta^2+q^2-\frac{\gamma^2 q^2}{\alpha^2+q^2}\right)\left|\phi_q\right|^2\right]\!,
\end{equation} which is found by completing the square. Here, $u_q$ and $\phi_q$ represent the Fourier transforms of $u(\mathbf{r})$ and $\phi(\mathbf{r})$, respectively. By neglecting the thermal fluctuations, we may integrate out one of the degrees of freedom \cite{leibler2}, which gives a new effective free-energy:
\begin{equation}
	\label{eqn:ftF-eff}
	\tilde{\mathcal{F}}_{q,0} = \frac{\gamma^2}{2\pi}\int\mathrm{d}^2\mathbf{q}\left[\frac{\alpha^2\beta^2 + \left(\alpha^2+\beta^2-\gamma^2\right)q^2+q^4}{\alpha^2+q^2}\right]\left|\phi_q\right|^2 =  \frac{\gamma^2}{2\pi}\int\mathrm{d}^2\mathbf{q}\,\frac{\left(k^2_{+}+q^2\right)\left(k^2_{-}+q^2\right)}{\alpha^2+q^2}\left|\phi_q\right|^2\!\!,
\end{equation} where the last step is obtained by using the definition of $k_\pm$ as given in (\ref{eqn:kSq-eqn}). The integrand of (\ref{eqn:ftF-eff}) becomes negative if and only if $k^2_\pm<0$ or, equivalently, $\gamma>\gamma_c=\alpha+\beta$, which corresponds to an instability with respect to the long wavelength fluctuations in membranes.

To understand the physical significance of the stable complex region of $k_\pm$, it is informative to consider a specific model of a membrane inclusion. Here, we choose to describe the inclusion as a circularly symmetric wedge-shaped transmembrane protein, where the height, $\mathcal{U}(\theta)$, and the contact angle, $\mathcal{U}'(\theta)$, are taken to be some constants $z_0$ and $\delta_0$, respectively. Using (\ref{eqn:final-u}), (\ref{eqn:final-phi}) and (\ref{eqn:z0}), the membrane profile $u(r)$ and the compositional asymmetry field $\phi(r)$ can be reduced to 
\begin{equation}
	\label{eqn:u-conical-incl}
	u(r) = \frac{r_0\,\delta_0}{f_+-f_-}\left[\frac{f_+\,K_0(k_-r)}{k_-r_0\,K_1(k_-r_0)} - \frac{f_-\,K_0(k_+ r)}{k_+r_0\,K_1(k_+ r_0)}\right]\!,
\end{equation} and 
\begin{equation}
	\label{eqn:phi-conical-incl}
	\phi(r) = r_0\,\delta_0\,\frac{f_+f_-}{f_+-f_-}\left[\frac{K_0(k_-r)}{k_-r_0\,K_1(k_-r_0)} - \frac{K_0(k_+ r)}{k_+r_0\,K_1(k_+ r_0)}\right]\!,
\end{equation} respectively. By approximating the modified Bessel functions by $K_n (\rho) \approx e^{-\rho}\sqrt{\frac{2}{\pi \rho}}$ for $\rho\gg n$ \cite{Abramowitz1965}, and by re-writing $k_\pm = \lambda \pm i \omega$, with 
\begin{equation}
	\label{eqn:depth-freq}
	\lambda = \frac{1}{2}\sqrt{(\alpha+\beta)^2 - \gamma^2}\quad\text{and}\quad \omega = \frac{1}{2}\sqrt{\gamma^2 - (\alpha-\beta)^2}
\end{equation} as real positive numbers, then after some algebra the asymptotic form of (\ref{eqn:u-conical-incl}) and (\ref{eqn:phi-conical-incl}), when $\gamma_d\leq\gamma<\gamma_c$, are found to be:
\begin{equation}
	\label{eqn:approx-u-conical-incl}
	u(r) \approx \frac{\gamma\,\mathfrak{C}_0\,e^{-\lambda(r-r_0)}}{\sqrt{r/r_0}}\,\cos\left[\,\omega\left(r-r_0\right) + \vartheta\,\right],\quad\text{and}\quad
	\phi(r) \approx \frac{\alpha\,\mathfrak{C}_0\,e^{-\lambda(r-r_0)}}{\sqrt{r/r_0}}\,\cos\left[\,\omega\left(r-r_0\right) + \zeta\,\right],
\end{equation} where
\begin{equation}
	\label{eqn:approx-conical-consts}
	\mathfrak{C}_0 = \frac{\delta_0}{2\,\omega\lambda}\sqrt{\alpha/\beta\,},
	\quad\vartheta = \arctan\left[\frac{\lambda\left(\beta\left(\alpha-\beta\right)+\gamma^2\right)}{\omega\left(\beta\left(\alpha+\beta\right)-\gamma^2\right)}\right]\!,\quad\text{and}\quad
	\zeta = \pi - \arctan\left(\lambda/\omega\right).
\end{equation} 

This shows that both $u(r)$ and $\phi(r)$ are no longer monotonically decreasing, displaying an underdamped behavior, where the wave-number $\omega$ of the radial profiles vanishes for a coupling term $\gamma = \gamma_d$, and their decay length, $1/\lambda$, diverges at $\gamma = \gamma_c$, which signals the presence of a Leibler-type instability. Although we restricted our analysis to a specific model, this behavior is true for lower symmetry inclusions as well, because the complex arguments of higher $K_n$ show similar underdamped decay.

\section{Dirichlet Boundary Condition}

Here, in addition to the boundary conditions on $u(\mathbf{r})$ as shown in (\ref{eqn:BC-u}) and (\ref{eqn:BC-u'}), we consider a Dirichlet boundary condition for $\phi(\mathbf{r})$ on $\partial\mathcal{M}$, namely
\begin{equation}
	\label{eqn:BC-phi-D}
	\phi(r_0,\theta) = \Phi(\theta)\equiv\sum_{n=0}^{\infty} \varphi_n \cos(n\theta - \nu_n),
\end{equation} where $\nu_0 = 0$. The function $\Phi(\theta)$ gives the compositional asymmetry field at the inference between the lipid bilayer and the rigid inclusion (of radius $r_0$), which together with the other boundary conditions completely determine the unknown functions $\mathcal{V}^\pm_n(\theta)$ and $\mathcal{W}_n(\theta)$ in (\ref{eqn:u}) and (\ref{eqn:phi}). As before, these conditions can written as matrix equation:
\begin{equation}
	\label{eqn:BC-matrix-D}
	\begin{pmatrix}
	\,K_n(k_+r_0)\, 			& \,K_n(k_-r_0)\, 			& \,1\, \\[1ex]
	\,\mathcal{L}^+_n K_n(k_+r_0)\, 	& \,\mathcal{L}^-_n K_n(k_-r_0)\, 	& \,n\, \\[1ex]
	\,f_+ K_n(k_+r_0)\, 			& \,f_- K_n(k_-r_0)\, 			& \,0\,
	\end{pmatrix}\begin{pmatrix}\mathcal{V}^+_n(\theta) \\[1ex] \mathcal{V}^-_n(\theta) \\[1ex] \mathcal{W}(\theta)\,r_0^{-n}\end{pmatrix} 
	= \begin{pmatrix}z_n\cos(n\theta-\xi_n) \\[1ex] r_0\,\delta_n\cos(n\theta-\chi_n) \\[1ex] \varphi_n\cos(n\theta-\nu_n)\end{pmatrix}\!,
\end{equation} with $n\geq0$ and $\mathcal{L}^\pm_n$ as defined in (\ref{eqn:Lpm}). By assuming that the matrix above is nonsingular, that is, 
\begin{equation}
	\label{eqn:P-n}
	\mathcal{P}_n \equiv f_-(n-\mathcal{L}^+_n)-f_+(n-\mathcal{L}^-_n) \neq 0,
\end{equation} this allows us to find that
\begin{equation}
	\label{eqn:Vn-D}
	 \mathcal{V}^{\pm}_{n}(\theta) = \pm\,\frac{\,f_\mp \mathcal{L}^{\mp}_{n}\left[n z_n \cos(n\theta - \xi_n)- r_0 \delta_n \cos(n\theta - \chi_n)\right] - (n-\mathcal{L}^{\mp}_n)\,\varphi_n\cos(n\theta-\nu_n)\,}{\mathcal{P}_{n}\,K_{n}(k_\pm r_0)}
\end{equation} and 
\begin{align}
	\label{eqn:Wn-D}
	\mathcal{W}_n(\theta) &= \frac{1}{r^{-n}_0\,\mathcal{P}_{n} }\left[z_n \left(f_+ \mathcal{L}_n^--f_- \mathcal{L}_n^+\right) \cos (n\theta-\xi_n) - \left(f_+-f_-\right) r_0\,\delta _n \cos (n\theta-\chi_n)\right.\notag \\ &\left.\qquad\qquad\qquad\qquad\qquad\qquad\qquad\qquad\qquad\qquad\qquad 
	+\,\left(\mathcal{L}_n^+-\mathcal{L}_n^-\right)\varphi_n\cos (n\theta-\nu_n)\right]\!.
\end{align}

Moreover, as the membrane profile $u(r,\theta)$ must vanishes in the far-field limit, then the height $z_0$ is found to be
\begin{equation}
	\label{eqn:z0-D}
	z_0 = \frac{\left(f_{+}-f_{-}\right)r_0\,\delta_0 - \left(\mathcal{L}_0^+-\mathcal{L}_0^-\right) \varphi_0}{f_+ \mathcal{L}_0^--f_- \mathcal{L}_0^+}.
\end{equation} 

Through the functions $u(r,\theta)$ and $\phi(r,\theta)$, the total deformation energy of the membrane, as given in equation (\ref{eqn:noFBCs0}), can be now computed as a Fourier series,
\begin{equation}
	\label{eqn:F-series-D}
	\tilde{\mathcal{F}} = \sum^{\infty}_{n=0} \tilde{\mathcal{F}}_n + \tilde{\mathcal{F}}_\text{tilt},
\end{equation} where $\tilde{\mathcal{F}}_n$ is the energy associated to each Fourier-Bessel mode, that is, \begin{align}
	\label{eqn:F-series-n-D}
	& \tilde{\mathcal{F}}_n 
	= \frac{\,\gamma^2 \varphi^2_n\,}{\mathcal{P}_n\,\vartheta_n} \left[f_{-}\mathcal{L}^{-}_n (n-\mathcal{L}^{+}_n) - f_{+}\mathcal{L}^{+}_n (n-\mathcal{L}^-_n)\right] + \frac{\alpha^2}{\mathcal{P}_n\,\vartheta_n} \left\{n\,z_n^2\,(f_{+} \mathcal{L}^{-}_n - f_{-}\mathcal{L}^{+}_n) 
	+ (f_{+} - f_{-})\,r_0^2\delta_n^2 \right. \notag\\[0.25ex]
	& \left. \quad - 2\,n\,z_n r_0 \delta_n\,(f_{+} - f_{-})\cos(\xi_n-\chi_n) + 2\,\varphi_n\,(\mathcal{L}_n^+-\mathcal{L}_n^-)\left[n z_n \cos (\nu_n-\xi _n) - r_0\delta _n \cos(\nu_n-\chi_n)\right]\right\}\!,
\end{align} with $\vartheta_0 = 1$ and $\vartheta_n = 2$ for any $n>0$. The second term in (\ref{eqn:F-series-D}) is an additional energetic contribution due to the balance of torques on the rigid inclusion, which requires that the total energy to be invariant under a tilt transformation of the form (\ref{eqn:balance}). Thus, the explicit form of $\tilde{\mathcal{F}}_\text{tilt}$ can be determined as follows:
\begin{align}
	\label{eqn:F-tilt-D}
	\tilde{\mathcal{F}}_\text{tilt} &= \frac{\alpha^2}{2\,\mathcal{P}_1}\,\left[f_-\mathcal{L}^-_1(3+\mathcal{L}^+_1) - f_+\mathcal{L}^+_1(3+\mathcal{L}^-_1)\right]\!\left(z^2_\text{tilt} - 2\,z_\text{tilt}\,r_0 \delta_\text{tilt} \cos(\xi_1-\chi_1)+r^2_0\delta^2_\text{tilt}\right.
	\notag \\
	& \qquad\qquad\qquad\left. - \,2\,\varphi_\text{tilt}\,r_0 \delta_\text{tilt} \cos(\nu_1-\chi_1) + 2\,z_\text{tilt}\,\varphi_\text{tilt}\cos(\xi_1-\nu_1) + \varphi^2_\text{tilt} \right)\!,
\end{align} where the coefficients $z_\text{tilt}$, $\delta_\text{tilt}$ and $\varphi_\text{tilt}$ are given by
\begin{equation}
	\delta_\text{tilt} = \delta_1\,\frac{2\left(f_+ - f_-\right)}{f_+\left(3+\mathcal{L}^-_1\right) - f_-\left(3+\mathcal{L}^+_1\right)},\quad
	z_\text{tilt} = z_1\,\frac{f_+\left(1+\mathcal{L}^-_1\right) - f_-\left(1+\mathcal{L}^+_1\right)}{f_+\left(3+\mathcal{L}^-_1\right) - f_-\left(3+\mathcal{L}^+_1\right)},\quad
\end{equation} and 
\begin{equation}
	\varphi_\text{tilt} = \varphi_1\,\frac{2\left(\mathcal{L}^+_1 - \mathcal{L}^-_1\right)}{f_+\left(3+\mathcal{L}^-_1\right) - f_-\left(3+\mathcal{L}^+_1\right)},
\end{equation} respectively. Moreover, this leads to the additional fields $u_\text{tilt}(r,\theta)$ and $\phi_\text{tilt}(r,\theta)$, which give the corresponding contributions to $u(r,\theta)$ and $\phi(r,\theta)$ that allow for a vanishing net torque. As shown in (\ref{eqn:u-tilt}) and (\ref{eqn:phi-tilt}), they can be written in terms of the azimuthal functions $\mathcal{V}^\pm_\text{tilt}(\theta)$ and $\mathcal{W}_\text{tilt}(\theta)$, which are found to be
\begin{equation}
	\label{eqn:V-tilt-D}
	\mathcal{V}^\pm_\text{tilt}(\theta) = \mp\,\frac{f_\mp\left[z_1 \cos(\theta - \xi_1) + r_0  \delta_1 \cos(\theta - \chi_1)\right] + \left(3 + \mathcal{L}^\mp_1\right) \varphi _1 \cos(\theta - \nu_1)}{\left[f_+\,(3+\mathcal{L}_1^-)-f_-\,(3+\mathcal{L}_1^+)\right] K_1(k_\pm r_0)},
\end{equation} and 
\begin{equation}
	\label{eqn:W-tilt-D}
	\mathcal{W}_\text{tilt}(\theta) = \frac{(f_+-f_-)\left[z_1 \cos(\theta - \xi_1) + r_0  \delta_1 \cos(\theta - \chi_1)\right] - \left(\mathcal{L}_1^+-\mathcal{L}_1^-\right) \varphi _1 \cos(\theta - \nu_1)}{\left[f_+\,(3+\mathcal{L}_1^-)-f_-\,(3+\mathcal{L}_1^+)\right] r^{-1}_0}.
\end{equation}

Interestingly, if the total deformation energy (\ref{eqn:F-series-D}) is minimized with respect to $\varphi_n$ and $\nu_n$, then we retrieve the same solutions as found for the Neumann boundary case. Therefore, in the absence of any constrains of the compositional asymmetry $\phi$ at the inclusion-membrane interface, the Neumann boundary condition gives the most energetically favorable membrane configurations.

\section{Deformation Profiles due to Protein Coats}

So far, we only examined the deformation profiles near a transmembrane inclusion, namely the mid-plane of the bilayer, $u(\mathbf{r})$, and its corresponding local compositional asymmetry, $\phi(\mathbf{r})$. Here, we consider the deformation fields induced by a protein coat \cite{molecular-bio} modeled as a spherical rigid object that adheres to the lipid bilayer. The membrane conformation on which the protein coat adheres is assumed to be commensurate with the shape of the coat, which has a constant intrinsic curvature $1/\mathcal{R}_c$. In the early stages of growth, the formation of this spherical cap (say, of radius $r_0$) only weakly perturbs the surrounding membrane, with a small contact angle $\delta_0 \approx r_0 / \mathcal{R}_c$. Consequently, the outer membrane can be described using the methodology developed in the previous sections, where the region of the membrane that covers the protein coat (which will denote here by $\mathcal{M}_c$) acts as a conical-shaped inclusion. Thus, we have that
\begin{equation}
	\label{eqn:coat-outer-u}
	u(r) = r_0\,\delta_0\left[\frac{f_+ \mathcal{L}^+_0 K_0(k_-r)}{\mathcal{P}_0\,K_0(k_-r_0)} - \frac{f_- \mathcal{L}^-_0 K_0(k_+r)}{\mathcal{P}_0\,K_0(k_+r_0)}\right] - \varphi_0 \left[\frac{\mathcal{L}^+_0 K_0(k_-r)}{\mathcal{P}_0\,K_0(k_-r_0)} - \frac{\mathcal{L}^-_0 K_0(k_+r)}{\mathcal{P}_0\,K_0(k_+r_0)}\right]\!,
\end{equation} and
\begin{equation}
	\label{eqn:coat-outer-phi}
	\phi(r) = r_0\,\delta_0\left[\frac{f_- f_+\mathcal{L}^+_0 K_0(k_-r)}{\mathcal{P}_0\,K_0(k_-r_0)} - \frac{\,f_- f_+\mathcal{L}^-_0 K_0(k_+r)}{\mathcal{P}_0\,K_0(k_+r_0)}\right] - \varphi_0 \left[\frac{f_- \mathcal{L}^+_0 K_0(k_-r)}{\mathcal{P}_0\,K_0(k_-r_0)} - \frac{f_+ \mathcal{L}^-_0 K_0(k_+r)}{\mathcal{P}_0\,K_0(k_+r_0)}\right]\!,
\end{equation} where radial distance $r\ge r_0$ and the Dirichlet boundary condition is used for the compositional asymmetry field, namely $\phi(r_0) = \varphi_0$. However, $\varphi_0$ is not completely arbitrary, being set as the value that minimizes the total membrane energy inside and outside of the protein coat. 

The deformation energy of the outer membrane, through equation (\ref{eqn:F-series-n-D}), is given by
\begin{equation}
	\label{eqn:enery-outer-coat}
	\tilde{\mathcal{F}}_\text{outside} = \frac{\alpha^2\left[\left(f_{+} - f_{-}\right)r_0^2\,\delta_0^2 - 2\,\varphi_0\left(\mathcal{L}_0^+-\mathcal{L}_0^-\right)r_0\delta_0\,\right] - \mathcal{L}^{-}_0\mathcal{L}^{+}_0\left(k^2_{+}-k^2_{-}\right)\varphi^2_0\,}
	 {f_+\mathcal{L}^-_0 - f_-\mathcal{L}^+_0}.
\end{equation}

As analogous to equation (\ref{eqn:rF}), the nondimensionalized free-energy functional of the membrane within the region $\mathcal{M}_c$  is found to be
\begin{equation}
	\label{eqn:energy-inside-coat}
	\tilde{\mathcal{F}}_\text{inside} = 4\left(1-\cos\delta_0\right) + 
	\alpha^2\,\mathcal{R}^2_c\left(1-\cos\delta_0\right)^2 + 
	\frac{\gamma^2}{2\pi}\int_{\mathcal{M}_c} \!\mathrm{d}^2\mathbf{r}\left[\beta^2\,\phi^2 + (\nabla \phi)^2 + \frac{4\,\phi\,}{\,\mathcal{R}_c\,}\right]\!,
\end{equation} which admits the following Euler-Lagrange equation:
\begin{equation}
	\label{eqn:EL-inside-coat}
	\left(\nabla^2 - \beta^2\right)\phi - 2/\mathcal{R}_c = 0.
\end{equation}
By assuming that compositional asymmetry field $\phi$ remains finite everywhere within $\mathcal{M}_c$ and using as before the same Dirichlet boundary condition, $\phi(r_0) = \varphi_0$, this differential equation yields that
\begin{equation}
	\label{eqn:sol-phi-inside-coat}
	\phi(r) = -\frac{2}{\beta^2\,\mathcal{R}_c} + \left(\varphi_0 + \frac{2}{\beta^2\,\mathcal{R}_c}\right)\frac{I_0(\beta\,r)}{I_0(\beta\,r_0)},
\end{equation} where $I_0$ is the modified Bessel function of the first kind of order zero. Therefore, the energy (\ref{eqn:energy-inside-coat}) to lowest order in $\delta_0$ becomes
\begin{equation}
	\label{eqn:min-energy-inside-coat}
	\tilde{\mathcal{F}}_\text{inside} = 2\,\delta_0^2 \left(1-\frac{\gamma^2}{\beta^2}\right) + \frac{r_0\,\gamma^2 \left(2 + \varphi_0\,\beta^2\,\mathcal{R}_c\right)^2 I_1(\beta\,r_0)}{\beta^3\,\mathcal{R}^2_c\,I_0(\beta\,r_0)},
\end{equation} with $I_1$ as the first order modified Bessel function of the first kind. The value of $\varphi_0$ is obtained by minimizing the total deformation energy $\tilde{\mathcal{F}}_\text{total}= \tilde{\mathcal{F}}_\text{inside} + \tilde{\mathcal{F}}_\text{outside}$, which gives that
\begin{equation}
	\label{eqn:varphi-0-coat}
	\varphi_0 = r_0\,\delta_0\frac{f_-f_+\left(\mathcal{L}_0^+-\mathcal{L}_0^-\right) \beta\,r_0\,I_0\left(\beta\,r_0\right) + 2\left(f_-\mathcal{L}_0^+-f_+\mathcal{L}_0^-\right)I_1\left(\beta\,r_0\right)}{\beta\,r_0\left[\mathcal{L}_0^- \mathcal{L}_0^+ (f_+-f_-) I_0\left(\beta\,r_0\right)+
	\left(f_+ \mathcal{L}_0^--f_- \mathcal{L}_0^+\right) \beta\,r_0\,I_1\left(\beta\,r_0\right)\right]},
\end{equation} where $r_0 = \delta_0\mathcal{R}_c$ is employed to simplify the expression. This allows us to determine the total deformation energy $\tilde{\mathcal{F}}_\text{total}$ due to a protein coat in the early stages of growth as a function of the inclusion radius $r_0$. Moreover, the membrane energy change due to coupling to $\phi$ only and scaled by the coat area, namely

\begin{equation}
	\Delta f_c  = \frac{1}{\displaystyle\pi r_0^{2}}\left(\tilde{\mathcal{F}}_\text{total} - \lim_{\gamma\,\to\,0}\tilde{\mathcal{F}}_\text{total}\right)\!,
\end{equation} can be computed to quadratic order in the angle $\delta_0$ (noting that $r_0$ and $\delta_0$ are not independent variables). This object renormalizes the chemical potential for binding of early coat monomers to the membrane and gives a measure of the energy landscape due to lipid compositional variation and its coupling to the mean curvature of the membrane.

\newpage